\documentclass{article}

\usepackage[amssymb]{SIunits}

\usepackage[centertags]{amsmath}
\usepackage{chicago,fancyhdr}
\usepackage{amssymb,amsfonts,eucal,oldgerm}
\usepackage[text={6in,8.25in},centering]{geometry}
\usepackage[colorlinks=true,citecolor=red,linkcolor=blue,urlcolor=blue,pdftex]{hyperref}

\newcommand{\skipline}[1][1]{\vspace*{#1\baselineskip}}

\newcommand{\coloneq}{\mathrel{\mathop:}=}
\newcommand{\half}{\frac{1}{2}}
\newcommand{\afourth}{\frac{1}{4}}

\newtheorem{theorem}{Theorem}[subsection]

\newtheorem{postulate}[theorem]{Postulate}

\newcommand{\resetcounters}{%
  \setcounter{equation}{0}%
  \setcounter{theorem}{0}%
  \setcounter{figure}{0}%
}

\newcommand{\resetsub}{%
  \resetcounters%
  \renewcommand{\thetheorem}{\arabic{section}.\arabic{subsection}%
    .\arabic{theorem}}%
  \renewcommand{\theequation}{\arabic{section}.\arabic{subsection}%
    .\arabic{equation}}%
  \renewcommand{\thefigure}{\arabic{section}.\arabic{subsection}%
    .\arabic{figure}}%
}

\renewcommand{\thetheorem}{\arabic{section}.\arabic{theorem}}%
\renewcommand{\theequation}{\arabic{section}.\arabic{equation}}%
\renewcommand{\thefigure}{\arabic{section}.\arabic{figure}}%

\title{Classical Black Holes Are Hot\thanks{This paper has been
    submitted to \emph{Philosophy of Science}, August, 2014.  I thank
    The Young Guns of the Spacetime Church of the Angle Brackets for
    helpful discussion on an earlier draft of this paper, and Ted
    Jacobson for insightful and critical discussion of my proposed
    Carnot-Geroch cycle for classical black holes.  I also thank
    audiences at the physics and philosophy departments of the
    University of Western Ontario, at the 2011 European Philosophy of
    Science Association, at the Munich Center for Mathematical
    Philosophy, at the NYU/Columbia/Rutgers Philosophy of Physics
    Group, at the 2014 British Society for Philosophy of Science, and
    most especially at the Southern California Group on the Philosophy
    of Physics, for invaluable feedback on earlier versions of this
    work.  I thank in particular Harvey Brown, Craig Callender, Robert
    Geroch, John Manchak, Wayne Myrvold, Chris Smeenk, and Jim
    Weatherall for many fruitful and enjoyable discussions of all the
    issues I treat here, and in particular I thank Harvey for his
    relentless, bearish, and, most of all, extraordinarily helpful
    asking of me, ``So what?''.  Finally, I thank Robert Wald for
    mercilessly pushing me over the course of several enjoyable days
    of conversation on the problem of possible violations of the
    Generalized Second Law (which I discuss in some detail in
    \S\ref{sec:probs-poss-resol-insghts-qs}), and I thank Bill Unruh
    for being amused by the show.}}

\author{Erik Curiel\thanks{\textbf{Author's address}: Munich Center
    for Mathematical Philosophy, Ludwig-Maximilians-Universit\"at,
    Ludwigstra{\ss}e 31, 80539 M\"unchen, Germany; \textbf{email}:
    \href{mailto:erik@strangebeautiful.com}
    {\texttt{erik@strangebeautiful.com}}}}

\begin{document}

\maketitle 

\tableofcontents

\newpage

\begin{quote}
  \begin{center}
    \textbf{\large ABSTRACT}
  \end{center}

  In the early 1970s it is was realized that there is a striking
  formal analogy between the Laws of black-hole mechanics and the Laws
  of classical thermodynamics.  Before the discovery of Hawking
  radiation, however, it was generally thought that the analogy was
  only formal, and did not reflect a deep connection between
  gravitational and thermodynamical phenomena.  It is still commonly
  held that the surface gravity of a stationary black hole can be
  construed as a true physical temperature and its area as a true
  entropy only when quantum effects are taken into account; in the
  context of classical general relativity alone, one cannot cogently
  construe them so.  Does the use of quantum field theory in curved
  spacetime offer the only hope for taking the analogy seriously?  I
  think the answer is `no'.  To attempt to justify that answer, I
  shall begin by arguing that the standard argument to the contrary is
  not physically well founded, and in any event begs the question.
  Looking at the various ways that the ideas of ``temperature'' and
  ``entropy'' enter classical thermodynamics then will suggest
  arguments that, I claim, show the analogy between classical
  black-hole mechanics and classical thermodynamics should be taken
  more seriously, without the need to rely on or invoke quantum
  mechanics.  In particular, I construct an analogue of a Carnot cycle
  in which a black hole ``couples'' with an ordinary thermodynamical
  system in such a way that its surface gravity plays the role of
  temperature and its area that of entropy.  Thus, the connection
  between classical general relativity and classical thermodynamics on
  their own is already deep and physically significant, independent of
  quantum mechanics.
\end{quote}

\section{Introduction}
\label{sec:intro}

I aim in this paper to clarify the status of the analogy between
black-hole mechanics restricted to general relativity on the one hand
(\emph{i}.\emph{e}., with no input from quantum field theory on curved
spacetime or from any other type of semi-classical calculation) and
classical thermodynamics on the other (``classical'' in the sense that
no quantum and no statistical considerations come into play).  Based
on the striking formal similarities of the respective mathematical
formul{\ae} of the Zeroth, First, Second and Third Laws of classical
thermodynamics and of the mechanics of black holes in stationary,
axisymmetric, asymptotically flat spacetimes, as I discuss in
\S\ref{sec:laws}, the best particular analogies seem to be: (1) that
between the surface gravity of a black hole as measured on its event
horizon and the temperature of a classical system; and (2) that
between surface area of the horizon and entropy.\footnote{Both the
  surface gravity and the surface area in question are defined with
  respect to the orbits of the Killing fields in virtue of which the
  spacetime is qualified as `stationary' and `axisymmetric'.  See
  \citeN[ch.~12]{wald-gr} for details.}  When it is also noted that
black holes, like ordinary thermodynamical systems, are characterized
by a small number of gross parameters independent of any details about
underlying microstructure, and that each version of the First Law
states a conservation principle for essentially the same quantity as
the other, \emph{viz}., mass-energy, it becomes tempting to surmise
that some deep or fundamental connection between black holes and
thermodynamics is being uncovered.  But is it of real physical
significance in some sense?

The conventional answer to this question is `no'.  Because classical
black holes seem to be perfect absorbers, they would seem to have a
temperature of absolute zero, even when they have non-zero surface
gravity.  It is only with the introduction of quantum considerations,
the standard account runs, in particular the derivation of Hawking
radiation, that one finds grounds for taking the analogy seriously.
And yet the startling and suggestive fact remains that one can derive
laws for black holes formally identical to those of classical
thermodynamical systems from the fundamental principles of general
relativity itself with no aid from quantum field theory in curved
spacetime.  Does the use of quantum field theory in curved spacetime
offer the only hope for taking the analogy seriously?  I think the
answer is `no'.  To attempt to justify that answer, I shall begin by
arguing in \S\ref{sec:standard-argument} that the standard argument to
the contrary is not physically well founded, and in any event begs the
question.  I will, therefore, in \S\ref{sec:temp-ent-class-thermo},
look at the various ways that the ideas of ``temperature'' and
``entropy'' enter classical thermodynamics, which will suggest
arguments that show the analogy between classical black-hole mechanics
and classical thermodynamics should be taken seriously indeed, without
the need to rely on or invoke quantum mechanics.  If this is correct,
then there may already be a deep connection between general relativity
and classical thermodynamics on their own, independent of quantum
mechanics.

My arguments in this paper, however, are not only negative.  I do
think that the connection between gravitational and thermodynamical
phenomena intimated by the formal equivalence of their respective Laws
is of real physical significance.  My strongest argument in favor of
this claim is the construction, in \S\ref{sec:taking-bhs-seriously},
of the analogue of a Carnot cycle with the heat sink provided by a
stationary black hole.  In the process, the black hole's surface
gravity and area play, respectively, the physical roles of temperature
and entropy of an ordinary heat sink in an ordinary Carnot cycle.  The
process also grounds the construction of an absolute temperature scale
that applies both to black holes and to ordinary classical
thermodynamical systems.  Finally, there follows from the construction
the existence of a universal constant with the physical dimension
needed to give surface gravity the physical dimension of temperature
and area the physical dimension of entropy.  To put the icing on the
cake, I also formulate the analogues of the Clausius and Kelvin
Postulates---the bases for the introduction of temperature and entropy
in classical thermodynamics---in the context of classical black hole
thermodynamics, and give arguments for them at least as strong as the
arguments for their analogues in classical thermodynamics, based on
the laws and properties of black holes.

If surface gravity and area couple to ordinary thermodynamical systems
in the same way as temperature and entropy, respectively, do, and if
they are introduced into the theory using the same constructions and
arguments, then there can be no grounds for denying that they
physically \emph{are} a real temperature and entropy.  To put it more
provocatively, if my claim is correct, then gravity on its own,
independent of its relation to the other three known fundamental
forces so successfully treated by quantum field theory, already is a
fundamentally thermodynamical phenomenon.\footnote{If one could show
  that the sorts of arguments I give here could be translated into the
  framework of Newtonian gravitational theory, that would provide even
  stronger support for this last claim.}  I want to stress,
nonetheless, that I do not consider quantum effects to be irrelevant
when considering possible relations between gravitational physics and
thermodynamics.  I want only to argue for the idea that the analogy
between the laws of classical thermodynamics and those of black hole
mechanics in classical general relativity is robust and deep in its
own right.

I conclude the paper, in \S\ref{sec:probs-poss-resol-insghts-qs}, with
a discussion of possible problems with my arguments and constructions,
some remarks on possible lessons my conclusions, if correct, may
yield, and some open questions.

Before diving in, I should perhaps say, by way of background, that I
am curious about this question in the first place in part because of
my curiousity about the larger question of the relation between
thermodynamical characteristics of a physical system and the
possibility of always being able to or indeed always being required to
find an underlying statistical interpretation of those thermodynamical
characteristics.  That the laws of black hole mechanics follow from
the fundamental theory itself (in this case, general relativity), and
are not as with classical thermodynamics an independent adjunct
connected to the underlying fundamental (Newtonian) theory through the
use of statistical devices, could suggest that thermodynamics is
itself more of the nature of a fundamental theory than has been
thought since the advent of statistical mechanics---or at least that
thermodynamical characteristics and quantities of physical systems may
be fundamental to them in some way analogous to that of other
fundamental characteristics and dynamical quantities, such as the
possession of a stress-energy tensor, for example, and its
satisfaction of some form of covariant conservation principle.  In a
similar vein, these sorts of results may also perhaps lend support to
the idea that general relativity is an effective field theory, and the
Einstein field equation only an equation of state, \emph{\`a la}
\citeN{jacobson-thermo-st-einstein-eqn-ste}, and perhaps
\citeN{bredberg-etal-navst-einst} and
\citeN{lysov-strominger-einst-navst}.  If that is true, then the
entire program of ``quantizing gravity'' may be misguided from the
start.  Yet another possibility, contrary to that just mentioned, is
that one may take my arguments as showing that the signature of
quantum gravity, in particular the traces of whatever statistical
quantities it may give us for making traditional sense of the
thermodynamical phenomena I discuss here, show up already in purely
classical, non-statistical theory.\footnote{I thank Fay Dowker for
  elucidating this possibility in a very helpful way in conversation.}
Finally, and I think most importantly, my arguments lend \emph{prima
  facie} support to projects (especially in cosmology) that want to
attribute entropy generically to ``gravitational degrees of freedom'',
as in the work of \citeN{clifton-et-gravl-ent-propos}, and as required
by Penrose's Conformal Curvature Hypothesis
\cite{penrose-sing-time-asym}.

I do not intend to investigate these larger issues here, however.  I
intend to investigate only the status of the analogy between the laws
of classical thermodynamics on the one hand and those of black-hole
mechanics in classical general relativity on the other.  I mention
these larger issues only to give some of my motivation for this work,
and to place it in the context of important work being done in many
branches of theoretical physics today.

There are other motivations behind this project as well.  Although
philosophers of physics have recently begun to work on issues arising
from proposals for theories of quantum gravity, some of which take as
their starting points the seemingly thermodynamical character of
gravitational phenomena as exemplified by the laws of black-hole
mechanics, almost no philosophical work has been done investigating
the nature of this seemingly thermodynamical character as revealed by
the structures of general relativity and of quantum field theory
formulated on curved, relativistic spacetimes.  Because general
relativity and quantum field theory are well entrenched, clearly and
rigorously articulated physical theories, I believe it behooves
philosophers to study them, if not before, at least in conjunction
with work done on quantum gravity.

\section{The Laws of Black-Hole Mechanics and the Laws of
  Thermodynamics}
\label{sec:laws}

Within the context of general relativity, one can derive laws
describing the behavior of black holes in stationary, asymptotically
flat spacetimes bearing a remarkable resemblance to the classical laws
of equilibrium thermodynamics.  I restrict attention to the
asymptotically flat case, because that is the simplest natural
analogue of an isolated system for black holes in general
relativity.\footnote{The generalization of the idea of a black hole
  and of the Four Laws to the non-asymptotically flat case by
  \citeN{hayward-genl-laws-bhdyns}, by the use of so-called dynamical
  trapping horizons, is of great interest, but to treat them would
  take us beyond the scope of this paper.  Also, I will not discuss
  the so-called Minus-First Law of \citeN{brown-uffink-minus-first};
  much work has been done to prove, or at least argue for, its
  correlate in black-hole mechanics (though not referred to as such in
  that literature), that perturbed black holes tend to settle down to
  equilibrium, and, in particular, that the sorts of perturbations I
  consider here do not destroy the event horizon.  There are now
  strong plausibility arguments in favor of it
  \cite{hollands-wald-stab-bhs-black-brns}, but its status in
  black-hole mechanics is still, to my mind, very much up for grabs,
  though, as a betting man, my money is on there being arguments for
  it at least as strong as for the Third Law (which, perhaps, is not
  to say very much).}  I restrict attention to stationary black holes
because those are the simplest natural analogue of an equilibrated
system for black holes in general relativity.

Now, for the laws themselves:\footnote{For proofs of the laws for
  black holes, see \citeN[ch.~12]{wald-gr}, \citeN{israel-3rdlaw-bh},
  \citeN{wald-qft-cst}, and
  \citeN{wald-gao-proc-1st-genl-2nd-charged-rot-bhs}.}
\begin{description} 
    \item[Zeroth Law] \hspace*{1em} \\
  \begin{description}
    \skipline[-1]
      \item[[Thermodynamics\hspace*{-.4em}]] The temperature $T$ is
    constant throughout a body in thermal equilibrium.\footnote{This
      is not the standard formulation of the thermodynamical Zeroth
      Law, which is ``If two systems are in thermal equilibrium with a
      third, then each is in thermal equilibrium with the other''.
      Because the formulation I use and the standard formulation are
      essentially equivalent when the systems at issue are assumed to
      be thermally homogeneous, as is the case for all the types of
      system my constructions rely on, and because I think this is a
      reasonable restriction when treating the Zeroth Law in any case,
      this is not a problem for my arguments.  Indeed, standard
      statements of the meaning of ``thermal equilibrium'' usually
      include the qualification that the system be thermally
      homogeneous, in the sense that the system contain no boundary
      with a permeability to heat flow different than that of the rest
      of the system.}
      \item[[Black Holes\hspace*{-.4em}]] The surface gravity $\kappa$
    is constant over the event horizon of a stationary black hole.
  \end{description}
    \item[First Law] \hspace*{1em} \\
  \begin{description}
    \skipline[-1]
      \item[[Thermodynamics\hspace*{-.4em}]] 
    \[
    \text{d}E = T\text{d}S + p\text{d}V + \Omega\text{d}J
    \]
    where $E$ is the total energy of the system, $T$ the temperature,
    $S$ the entropy, $p$ the pressure, $V$ the volume, $\Omega$ the
    rotational velocity and $J$ the angular
    momentum.\footnote{Strictly speaking, this is not the First Law,
      but rather the Gibbs Relation, which is equivalent to the First
      Law for thermodynamical systems in equilibrium and for systems
      that deviate from equilibrium only ``quasi-statically''.  Since
      all my arguments involve only systems in equilibrium, and, as is
      standard in thermodynamical arguments, systems that deviate from
      equilibrium only by quasi-stationary effects, this is not a
      problem.}
      \item[[Black Holes\hspace*{-.4em}]]
    \[
    \delta M = \frac{1}{8\pi} \kappa\delta A + \Omega_\textsc{bh}
    \delta J_\textsc{bh}
    \]
    where $M$ is the total black hole mass, $A$ the surface area of
    its horizon, $\Omega_\textsc{bh}$ the ``rotational velocity'' of
    its horizon,\footnote{See \citeN[ch.~12, \S3,
      pp.~319--320]{wald-gr}.}  $J_\textsc{bh}$ its total angular
    momentum, and `$\delta$' denotes the result of a first-order,
    linear perturbation of the spacetime.\footnote{For an exact
      definition and thorough discussion of the perturbations used,
      see \citeN{wald-gao-proc-1st-genl-2nd-charged-rot-bhs}.  There
      is an oddity about this formulation of the law, however, that I
      have not seen addressed in the literature but is surely worth
      puzzling over.  While the $\delta$ acting on $M$ is the same as
      that acting on $J_\textsc{bh}$, it is not the same as that
      acting on $A$.  The $\delta$ acting on $M$ and $J_\textsc{bh}$
      represents a perturbation of a quantity taken asymptotically at
      spatial infinity; the other represents perturbations taken ``at
      the event horizon''.  I know of no other physically significant
      equation where different differential operators act on different
      mathematical spaces in such a way that, as in this case, there's
      no natural mapping between them.  What's going on here?}
  \end{description}
    \item[Second Law] \hspace*{1em} \\
  \begin{description}
    \skipline[-1]
      \item[[Thermodynamics\hspace*{-.4em}]] $\delta S \ge 0$ for any
    process in an isolated system.\footnote{Again, this is not the
      usual formulation of the Second Law in classical thermodynamics,
      which is standardly given as the Clausius or Kelvin Postulate
      (\emph{e}.\emph{g}., \citeNP[\S7]{fermi-thermo}).  Because the
      principle of entropy increase follows from either Postulate
      \cite[\S\S11--13]{fermi-thermo}, and because the appropriate
      analogues for those Postulates hold for black holes
      (\S\ref{sec:bh-clausius-kelvin} below), this again is not a
      problem for my arguments.}
      \item[[Black Holes\hspace*{-.4em}]] $\delta A \ge 0$ in any
    process.\footnote{Note that, because we are considering by fiat
      only asymptotically flat black holes, the appropriate analogue
      of an isolated classical thermodynamical system, it would be
      redundant to stipulate in the statement of the Law that the
      process takes place in an isolated system.  Indeed, the Area
      Theorem (as the Second Law for black holes is often called) is a
      a result in pure differential geometry, the only input with a
      physical interpretation required being the so-called null energy
      condition.  That condition essentially rules out only
      macroscopic fluxes of negative energy, so the scope of the
      quantifier in ``any process'' in the statement of the Law should
      be taken very broadly indeed.  In particular, one need not even
      assume the process is quasi-static, nor even that the processes
      are restricted to the sorts of first-order, linear perturbations
      used in the formulation and proof of the First Law.  (See
      \citeNP{curiel-primer-econds} for a discussion of the physical
      content of the null energy condition and its role in the proofs
      of the Laws for black holes.)}
  \end{description}

  \newpage

    \item[Third Law] \hspace*{1em} \\
  \begin{description}
    \skipline[-1]
      \item[[Thermodynamics\hspace*{-.4em}]] $T = 0$ is not achievable
    by any process.\footnote{I actually think this is a defective
      statement of the Third Law of thermodynamics.  (See,
      \emph{e}.\emph{g}., \citeNP{schrodinger-stat-therm},
      \citeNP{aizenman-lieb-3rd-law-degen-grnd-state} and
      \citeNP{wald-nernst-bh-thermo} for a discussion of some of its
      problems.)  \citeN{schrodinger-stat-therm} provides a far more
      satisfactory statement of the Third Law, which I think carries
      over well into black-hole thermodynamics.  I do not have room to
      go into the matter here, though.}
      \item[[Black Holes\hspace*{-.4em}]] $\kappa = 0$ is not
    achievable by any process.
  \end{description}
\end{description}

The most striking architectonic similarity between the
characterization of ordinary thermodynamical systems (in equilibrium)
by the laws of thermodynamics and the characterization of black holes
(in equilibrium, \emph{i}.\emph{e}., stationary) is that in each case
the behavior of the system, irrespective of any idiosyncracies in the
system's constitution or dynamical history, is entirely captured by
the values of a small number of physical quantities, 6 for ordinary
thermodynamical systems, 4 for black holes: in the former case, they
are temperature, entropy, pressure, volume, angular velocity and
angular momentum;\footnote{Of course, the First Law guarantees that
  not all these quantities will be independent, and, if one is
  considering a particular species of thermodynamical system, then one
  may have available an equation of state that will further reduce the
  number of independent quantities, but all that is beside the point
  for my purposes.}  in the latter, they are surface gravity, area,
angular velocity and angular momentum.  The Zeroth and Third Laws
suggest that we take the surface gravity of a black hole as the
analogue of temperature.  The Second Laws suggest that we take area as
the analogue of entropy.  This is consistent with the First Law, if we
treat $\frac{1}{8\pi} \kappa \delta A$ as the Gibbsian ``heat'' term
for a system in thermal equilibrium.  Indeed, if we do so then the
analogy for the First Law becomes exact: relativistically, energy just
is mass, so the lefthand side terms of the First Law for ordinary
systems and for black holes are not just analogous, they are
physically identical; likewise, $\Omega_\textsc{bh} \delta
J_\textsc{bh}$ as a work term in the law for black holes is physically
identical to the corresponding term in the law for ordinary systems.

Now the force of the question motivating this paper should be clear:
the mathematical analogy is perfect, and there are already some
indications that the analogy may reach down to the level of physics,
not just mathematics.  But how far should we take the analogy?  What
can it mean to take seriously the idea that the surface gravity of a
black hole is a physical temperature, and its area a physical entropy?

\section{The Standard Argument Does Not Work}
\label{sec:standard-argument}

There are well-known difficulties with taking the surface gravity of a
classical black hole to represent a physical temperature.  One
important method for defining the thermodynamic temperature of an
object derives from the theory of thermal radiation from black bodies.
If a normal black body immersed in a bath of thermal radiation settles
down to thermal equilibrium, it will itself emit thermal radiation
with a power spectrum characteristic of its equilibrium temperature as
measured using a gas thermometer.  This power spectrum can then be
used to define a temperature scale.  It is this definition of
thermodynamic temperature that is almost always (at times implicitly)
invoked when the claim is made that if one considers classical general
relativity alone then black holes, being perfect absorbers and perfect
non-emitters, have an effective temperature of absolute
zero.\footnote{See for example the remarks in
  \citeN{bardeen-carter-hawking73}, \citeN{carter73} and
  \citeN{wald99}.  There is another form of argument for attributing
  the temperature absolute zero to all classical black holes, that it
  seems to be possible to use them to convert heat into work with
  100\% efficiency.  I address this type of argument in
  \S\ref{sec:probs-poss-resol-insghts-qs}.}

To try to be a little more precise, I will offer a reconstruction of
the standard argument.  It is not given in exactly this form by anyone
else in the literature, but I think it captures both the spirit and
the letter of the orthodox view.  Put a Kerr black hole in a box with
perfectly reflective sides, which are far from the event horizon (in
the sense that they are many times farther away from the event horizon
``in natural spacelike directions'' than its own ``natural''
diameter).  Pervade the box with thermal radiation.  According to
classical general relativity, the black hole will absorb all incident
thermal radiation, and emit none, until eventually all thermal
radiation in the box (outside the event horizon) has vanished, so the
black hole must have a temperature of absolute zero.  Thus, the
surface gravity $\kappa$, which is never zero for a non-extremal Kerr
black hole, cannot represent a physical temperature of the black hole
in classical general relativity.  Conventional wisdom holds, as a
result, that if the formal similarities mentioned above were all there
were to the matter then they would most likely represent a merely
accidental resemblance or perhaps would indicate at best a superficial
relationship between thermodynamics and black holes, but in any event
would not represent the laws of classical thermodynamics as extended
into the realm of black holes.\footnote{The remarks of
  \citeN[p.~337]{wald-gr}, for example, are exemplary in this regard.}

In 1974, using semi-classical approximation techniques Hawking
discovered that stationary, axisymmetric black holes appear to radiate
as though they were perfect black-body emitters in thermal equilibrium
with temperature $\displaystyle \frac{\hbar}{2 \pi} \kappa$, when
quantum particle-creation effects near the black hole horizon are
taken into account \cite{hawking-bh-explode,hawking-part-create-bh}.
It is this result that is generally taken to justify the view that the
resemblances between the laws of black hole mechanics and the laws of
classical thermodynamics point to a fundamental and deep connection
among general relativity, quantum field theory and thermodynamics, and
in particular that $\kappa$ \emph{does} in fact represent the physical
temperature of a black hole, and therefore $A$ its
entropy.\footnote{See again, for example, the remarks of
  \citeN[p.~337]{wald-gr}.  Indeed, some of the most important
  researchers in the field make even stronger claims.
  \citeN[p.~944]{unruh-wald-acc-rad-gsl}, for example, claim that
  ``the existence of acceleration radiation [outside the event
  horizon, a fundamentally quantum phenomenon,] is vital for the
  self-consistency of black-hole thermodynamics.''}

I have two problems with this orthodoxy.  First, I find the physical
content of the standard argument not to stand up to scrutiny.  While
it is true that the Kerr black hole in the box, according to classical
general relativity, will emit no blackbody radiation while it absorbs
any incident on it, that is not the end of the story.  Classical
general relativity does tell us that the Kerr black hole will emit
some radiation, \emph{viz}., \emph{gravitational} radiation, while it
is perturbed by the infalling thermal radiation, and that
gravitational radiation will in fact couple with the thermal radiation
still outside the black hole.  If we are trying to figure out whether
purely gravitational objects, such as black holes, have
thermodynamical properties, we should surely allow for the possibility
that gravitational radiation, or, indeed, the exchange of
``gravitational energy'' in any form, may count as a medium for
thermodynamical coupling.\footnote{I use scare-quotes for
  `gravitational energy' because that is an infamously vexed notion in
  classical general relativity, with no cogent way known to localize
  it, and indeed strong reasons to think there can be no localization
  of it in general.  (See, \emph{e}.\emph{g}.,
  \citeNP{curiel-geom-objs-nonexist-sab-uniq-efe}.)  I will discuss
  this issue, and the potential problems it may raise for my
  arguments, in \S\ref{sec:probs-poss-resol-insghts-qs}.}  Indeed,
just as electromagnetic radiation turned out to be a medium capable of
supporting a physically significant coupling of electromagnetic
systems with classical thermodynamical systems, it seems \emph{prima
  facie} plausible that gravitational radiation may play the same role
for gravitational systems.  Just as ``heat'' for an electromagnetic
system may be measured by electromagnetic radiation, at least when
transfer processes are at issue, so it may be that ``heat'' for a
gravitational system may be measured by gravitational radiation, or
any form of exchange of gravitational energy, again at least when
transfer processes are at issue.  Electromagnetic energy is just not
the relevant quantity to track when analyzing the thermodynamic
character of purely gravitational systems.

Second, I do not think this definition of temperature is the
appropriate one to use in the context of a purely classical
description of black holes, for the electromagnetically radiative
thermal equilibrium of systems immersed in a bath of thermal radiation
is essentially a \emph{quantum} and \emph{statistical} phenomenon, by
which I mean one that can be correctly modeled only by using the
hypothesis that radiative thermal energy is exchanged in discrete
quanta and then computed correctly only with the use of statistical
methods.  To use that characterization of temperature to argue that we
must use quantum mechanics in order to take surface gravity seriously
as a physical temperature, therefore, is to beg the question.  If my
qualm is well founded, it follows that the standard argument does not
bear on the strength of the analogy as indicating a real physical
connection between classical general relativity and thermodynamics.
After all, if one is trying to determine the status of the analogy
between \emph{classical} gravitational theory and \emph{classical}
thermodynamics independently of any quantum considerations, then the
most appropriate characterizations of temperature to use are those
grounded strictly in classical thermodynamics itself.  (I make the
idea of this qualm precise in \S\ref{sec:probs-poss-resol-insghts-qs},
in discussing possible problems with my arguments.)

There is yet another \emph{prima facie} problem, however, with trying
to interpret surface gravity as a true temperature and area as a true
entropy, which my arguments so far do not address: neither has the
proper physical dimension.  In geometrized units, the physical
dimension of temperature is mass (energy), and entropy is a pure
scalar.  The physical dimension of surface gravity, however, is
mass$^{-1}$, and that of area mass$^2$.  There are no purely classical
universal constants, moreover, available to fix the dimensions by
multiplication or division.\footnote{All the classical universal
  constants, such as the speed of light and Newton's gravitational
  constant, are dimensionless.  This is actually a puzzling state of
  affairs, that surely deserves investigation.}  The only available
universal constant to do the job seems to be $\hbar$, which has the
dimension mass$^2$.\footnote{I am grateful to Ted Jacobson and Carlo
  Rovelli for pushing me on the issue of the physical dimensions of
  the quantities, and on the seeming need to introduce $\hbar$ to make
  things work out properly.}  I cannot address this problem at this
stage of my arguments.  Remarkably, however, it will turn out as a
natural sequela to my construction of the appropriate analogue of a
Carnot cycle for black holes, in \S\ref{sec:schwarz-bh-carnot-cycles},
that the existence of a universal constant in the classical regime
with the proper dimension is guaranteed.

\section{Temperature and Entropy in Classical Thermodynamics}
\label{sec:temp-ent-class-thermo}

I think there are grounds for taking the analogy very seriously even
when one restricts oneself to the classical theories, without input
from or reliance on quantum theories.  To make the case more poignant,
imagine that we are physicists who know only classical general
relativity and classical thermodynamics, but have no knowledge of
quantum theory.  How could we determine whether or not to take black
holes as thermodynamical objects in a substantive, physical sense,
given that we know the deep formal analogy between the two sets of
laws?  In such a case, we ought to look to the way that temperature
and entropy are introduced in classical thermodynamics and the various
physical roles they play there.  If the surface gravity and area of
black holes can be introduced in the analogous ways and play the
analogous physical roles, I contend that the global analogy is already
on strong ground.  In other words, the surface gravity and area must
play the same role in the new theory \emph{vis}-\emph{\`a}-\emph{vis}
other theoretical quantities as temperature and entropy do in the
original theory \emph{vis}-\emph{\`a}-\emph{vis} the analogous
theoretical quantities there.  If, moreover, it can be shown that
surface gravity couples to ordinary classical thermodynamical systems
in the same formal way as ordinary temperature does, then there are no
grounds for denying that it is a true physical
temperature.\footnote{Since entropy directly mediates no coupling
  between thermodynamical systems, the same argument is not available
  for it.  This is one of the properties of entropy that makes it a
  truly puzzling physical quantity: there is no such thing, not even
  in principle, as an entropometer.}  And if area for black holes is
related to surface gravity and to the proper analogue of heat in the
same way as entropy is to ordinary temperature and heat, and if it is
required for formulating an appropriately generalized Second Law, then
there are no grounds for denying that it is a true physical entropy.
Indeed, it was exactly on grounds such as these that physicists in the
19th century concluded that the power spectrum of blackbody radiation
itself encoded a \emph{physical} temperature and entropy, not merely
that there was an analogy between thermodynamics and the theory of
blackbody radiation.  \citeN{planck97/26} himself had doubts about the
thermodynamical character of blackbody radiation until he had
satisfied himself on these points.

There are three fundamental, related ways that temperature is
introduced in classical thermodynamics, which themselves ground the
various physical roles temperature can play in the theory (how it
serves as the mediator of particular forms of coupling between
different types of physical system, \emph{e}.\emph{g}.).  The first
derives from perhaps the most basic of the thermodynamic
characteristics of temperature and is perhaps most definitive of the
cluster of ideas surrounding the concepts of ``temperature'' and
``heat'': it is that when two bodies are brought into contact, heat
will spontaneously flow from the one of higher temperature to the one
of lower temperature.\footnote{It is important for some of my later
  arguments to note that this characterization of comparative
  temperature does not preclude processes in which heat at the same
  time flows from the colder body to the hotter.  It says only that it
  is always the case that heat flows from hotter to colder,
  irrespective of what may or may not happen in the reverse
  direction.}  The second arises from the fact that increase in
temperature is positively correlated with increases in the capacity of
a system to do work.\footnote{See, \emph{e}.\emph{g}., the exemplary
  remarks of \citeN[p.~36]{sommerfeld-thermo}: ``Thermodynamics
  investigates the conditions that govern the transformation of heat
  into work. It teaches us to recognize temperature as the measure of
  the work-value of heat. Heat of higher temperature is richer, is
  capable of doing more work.''}  This fact allows one to define an
empirical scale of temperature through, \emph{e}.\emph{g}., the use of
a gas thermometer: the temperature reading of the thermometer is made
directly proportional to the volume of the thermometric gas used,
which is itself directly proportional to the work the gas does on its
surrounding container as it expands or contracts in response to its
coupling with the temperature of the system being measured.  The
utility of such a scale is underwritten by the empirical verification
that such empirical scales defined using a multitude of different
gases under a multitude of different conditions are consistent among
one another.\footnote{\citeN[{\S}1, p.~1]{planck97/26} remarks that
  quantitative exactness is introduced into thermodynamics through
  this observation, for changes of volume admit of exact measurements,
  whereas sensations of heat and cold do not, nor even comparative
  judgments of hotter and cooler on their own.}  The third arises from
an investigation of the efficiency of reversible, cyclic engines,
\emph{viz}., Carnot engines, which yields a definition of the
so-called absolute temperature scale associated with the name of
Kelvin.\footnote{See, \emph{e}.\emph{g}.,
  \citeN[\S\S8--10]{fermi-thermo}.}  It is the possibility of
physically identifying the formally derived absolute scale with the
empirically derived scale based on capacity to do work (increase in
volumes, \emph{e}.\emph{g}.) that warrants the assertion that they
both measure the same physical
quantity.\footnote{\citeN[chs.~\textsc{viii},
  \textsc{xiii}]{maxwell-theory-heat-1888} gives a wonderfully
  illuminating discussion of the physical basis of the equivalence of
  the absolute temperature scale with the one based on gas
  thermometry.}

Likewise, there are (at least) three ways that entropy enters
classical thermodynamics.  The first historically, and perhaps the
most physically basic and intuitive, is as a measure of how much
energy it takes to transform the heat of a thermal system into work:
generally speaking, the free energy of a thermodynamical system is
inversely proportional to its entropy.\footnote{Again, the discussion
  of \citeN[ch.~\textsc{xii}]{maxwell-theory-heat-1888} about this
  idea is a masterpiece of physical clarification and insight.}  The
second is as that perfect differential $\text{d} S$ into which
temperature, as integrating factor, transforms exchanges of heat
$\text{d} Q$ over the course of quasi-stationary processes
\cite[ch.~\textsc{iv}]{fermi-thermo}: the integral of $\text{d} Q$
along a quasi-stationary path between two equilibrium states in the
space of states of a thermodynamical system is not independent of the
path chosen, whereas the integral of $\displaystyle \frac{\text{d}
  Q}{T}$ is.  (Indeed, \citeNP{sommerfeld-thermo} uses this fact to
conclude that entropy is a true physical property of a thermodynamical
system, whereas heat content is not.)  The third also arises from the
analysis of the efficiency of Carnot cycles
\cite[ch.~\textsc{iv}]{fermi-thermo}.

Now, the following fundamental theorem of classical thermodynamics
provides the basis both for the definition of the absolute temperature
scale and for the introduction of entropy as the perfect differential
derived from exchanges of heat when that temperature is used as an
integrating factor.
\begin{theorem}
  \label{thm:revers-eng-eff}
  Any two reversible, cyclic engines operating between temperatures
  $T_{2}$ and $T_{1}$ (as measured using gas thermometry) have the
  same efficiency.  The efficiency of any non-reversible engine
  operating between $T_{2}$ and $T_{1}$ is always less than this.
\end{theorem}
This theorem is a direct consequence of either the Clausius or the
Kelvin postulate, which can be argued on physical grounds both to be
equivalent to each other and to directly imply the principle of
entropy increase (for the proofs of which statements see,
\emph{e}.\emph{g}., \citeNP{fermi-thermo}):
\begin{postulate}[Lord Kelvin]
  \label{post:kelvin}
  A transformation whose only final result is to transform into work
  heat extracted from a source that is at the same temperature
  throughout is impossible.  
\end{postulate}
\begin{postulate}[Clausius]
  \label{post:clausius}  
  A transformation whose only final result is to transfer heat from a
  body at a given temperature to a body of a higher temperature is
  impossible.  
\end{postulate}
I claim that these last two postulates, and the fact that they provide
grounds for the proof of the efficiency theorem, for the introduction
of temperature and entropy as physical quantities, and for proof of
the principle of entropy increase, encode essentially all that is of
physical significance in the ways I sketched that both temperature and
entropy enter into classical thermodynamics.

The Clausius Postulate captures the idea that when two bodies are
brought into thermal contact, heat flows from the body of higher
temperature to the other.  The Kelvin Postulate captures the idea that
the capacity of a body to do work on its environment tends to increase
as its temperature increases.  If one could show that appropriately
formulated analogues to these two propositions about classical black
holes hold in general relativity, with surface gravity playing the
role of temperature and area that of entropy, one would have gone a
long way towards showing that surface gravity \emph{is} a true
thermodynamical temperature and area a true entropy.  If one could
further show that ordinary thermodynamical systems equilibrate with
black holes in a way properly mediated by their ordinary temperature
and by the black hole's surface gravity, so as to allow for the
construction of a Carnot-like cycle and the definition of an absolute
temperature scale, the analogy would have been shown to be far more
than analogy: it would be physical equivalence in the strongest
possible sense.  I prove all these propositions in
\S\ref{sec:taking-bhs-seriously} below.

\section{Taking Black Holes Seriously as Thermodynamical Objects}
\label{sec:taking-bhs-seriously}

\resetsub

What is needed, first, is a way to characterize ``thermal coupling''
between black holes and ordinary thermodynamical systems: granted that
``heat'' in the gravitational context is gravitational energy of a
particular form, such as that carried in the form of gravitational
radiation or that responsible for red-shift effects in monopole
solutions, then it follows that black holes \emph{are not perfect
  absorbers}.  When there is an ambient electromagnetic field, the
black hole will radiate gravitationally as it absorbs energy and grows
from the infalling electromagnetic radiation.  So to conclude that
surface gravity is a physical temperature, one need show only that the
gravitational energy exchanged between a black hole and other
thermodynamical systems in transfer processes depends in the
appropriate way on the surface gravity of the event
horizon.\footnote{I will discuss in
  \S\ref{sec:probs-poss-resol-insghts-qs} below the fact that there is
  no well defined notion of localized gravitational energy in general
  relativity, and how that may bear on my arguments.}  This approach
has \emph{prima facie} physical plausibility: to take the energy in
gravitational radiation, \emph{e}.\emph{g}., to be the gravitational
equivalent of heat is the same as to take the energy in
electromagnetic radiation to be the electromagnetic equivalent of
heat---it is what couples in the appropriate way to the average
kinetic energy of molecules in ordinary thermodynamical systems,
\emph{viz}., what makes it increase and decrease, and that with
respect to which equilibrium is defined.

Just as the concept of ``thermal coupling'' had to be emended in the
extension of classical thermodynamics to include phenomena associated
with radiating black bodies, so we should expect it to be in this
case.  In classical thermodynamics before the inclusion of black-body
phenomena, thermal coupling meant immediate spatial contiguity: heat
was known to flow among solids, liquids and gases only when they had
surfaces touching each other.\footnote{This fact, perhaps, contributed
  to the historical idea that heat was a fluxional, perhaps even
  fluid, substance, such as phlogiston or caloric.}  In order to
extend classical thermodynamics to include black-body phenomena, the
idea of thermal coupling had to be extended as well: two black bodies
thermally couple when and only when the ambient electromagnetic field
each is immersed in includes direct contributions from the
electromagnetic radiation emitted by the other.  They do not need to
have surfaces touching each other.

In order to characterize the correct notion of thermal coupling among
systems including black holes (or more generalized purely
gravitational systems, such as cosmological horizons), we first need
to characterize an appropriate notion of ``heat'' for black holes, and
the concomitant notion of free energy.  That will put us in a position
to construct the appropriate generalization of Carnot cycles for them,
and so to formulate the appropriate generalizations of the Clausius
and Kelvin Postulates for such systems.

\subsection{Irreducible Mass, Free Energy and ``Heat'' of Black
  Holes}
\label{sec:irred-mass-free-energy-heat}

In analyzing the ideas of reversibility and irreversibility for
processes involving black holes, \citeN{christodoulou70} introduced
the \emph{irreducible mass} $M_\text{irr}$ of a black hole of mass $M$
and angular momentum $J$:\footnote{I will discuss only Kerr black
  holes, not Kerr-Newman black holes that also have electric charge,
  as the ensuing technical complications would not be compensated by
  any gain in physical comprehension.}
\[
M^2_\text{irr} \coloneq \half [M^2 + (M^4 - J^2)^{\half}]
\]
(From hereon, I shall drop the subscripted `$\textsc{bh}$' on terms
denoting quantities associated with black holes, except in cases where
ambiguity may arise.)  Inverting the definition yields
\[
M^2 = M^2_\text{irr} + \afourth \frac{J^2}{M^2_\text{irr}}
\]
and so, for a Kerr black hole, 
\[
M > M_\text{irr}
\]
(Clearly, $M_\text{irr} = M$ for a Schwarzschild black hole.)  Thus,
the initial total mass of a black hole cannot be reduced below the
initial value of $M_\text{irr}$ by any physical process.  A simple
calculation for a Kerr black hole, moreover, shows that,
\begin{equation}
  \label{eq:area-irr-mass}
  A = 16\pi M^2_\text{irr}
\end{equation}
Thus, it follows from the Second Law that $M_\text{irr}$ itself cannot
be reduced by any physical process, and so any process in which the
irreducible mass increases is a physically irreversible process.  In
principle, therefore, the free energy of a black hole is just $M -
M_\text{irr}$, in so far as its total mass $M$ represents the sum
total of all forms of its energies, and $M_\text{irr}$ represents the
minimum total energy the black hole can be reduced
to.\footnote{Some---\emph{e}.\emph{g}., \citeN[ch.~12,
  \S4]{wald-gr}---interpret $M - M_\text{irr}$ as the rotational
  energy of a Kerr black hole, in so far as extracting that much
  energy from a black hole would necessarily reduce its angular
  momentum to zero.  Based on the arguments I will give in this
  section, I prefer to think of it as a thermodynamical free energy,
  which cannot necessarily be decomposed in a canonical way into
  different ``forms'', \emph{e}.\emph{g}., that much heat and that
  much rotational energy, \emph{etc}.}

In classical thermodynamics, it makes no sense to inquire after the
absolute value of the quantity of heat a given system possesses.  In
general, that is not a well defined property accruing to a system.
One rather can ask only about the amount of heat tranferred between
bodies during a given process.\footnote{\label{fn:max-heat}See
  \citeN[chs.~\textsc{i, iii, iv, viii,
    xii}]{maxwell-theory-heat-1888}.}  Consider, then, a classical
thermodynamical system with total energy $E$ and free energy
$E_\text{f}$.  $E - E_\text{f}$ is the amount of energy unavailable
for extraction, what Kelvin called its dissipated energy,
$E_\text{d}$.  Say that through some quasi-stationary process, we know
not what, both $E$ and $E_\text{d}$ change so that the system now has
less free energy than it did before; therefore, the entropy of the
system must have increased, which can happen only when it absorbs
heat, which will in general be the difference between the total change
in energy and the change in free energy.  If they both change so that
the system has more free energy, the same reasoning applies, and it
must have given up a quantity of heat equal to that difference.

These remarks suggest defining the ``quantity of heat transferred'' to
or from a black hole during any quasi-stationary thermodynamical
process to be the change in its free energy, which is to say the
change in total black hole mass minus the change in its irreducible
mass, $\Delta M - \Delta M_\text{irr}$.\footnote{I thank Harvey Brown
  for drawing to my attention the fact that
  \citeN{caratheodory-grund-thermo}, in his ground-breaking
  axiomatization of classical thermodynamics, introduced the notion of
  heat in a way very similar to this, not as a primitive quantity as
  is usually done, but as the difference between the internal and the
  free energies of a system.}  If, for instance, the irreducible mass
of a black hole does not change, while the total mass decreases, then
it would have given up a quantity of heat.  As a consistency check, it
is easy to see that, according to this definition, when an ordinary
thermodynamical system in equilibrium is dumped into a Kerr black
hole, the black hole absorbs the quantity of heat the ordinary matter
contained as characterized by the Gibbs relation, \emph{viz}., its
temperature times its entropy, as only that energy contributes to its
total mass without directly changing its angular momentum.  Based on
this characterization of ``quantity of heat transferred'', I claim
that the appropriate notion of thermal coupling for systems involving
black holes is any interaction where there is a change in the black
hole's free energy.  For purely gravitational interactions, this
includes emission and absorption of that part of the energy of
gravitational radiation not due to angular momentum, energy exchange
due to simple monopole- or multipole-moment couplings in the
near-stationary case, and so on.

Some care must be taken in applying this definition to Schwarzschild
black holes, however.  Because $M = M_\text{irr}$ for a Schwarzschild
black hole, one can never give up heat while remaining
Schwarzschildian.  Schwarzschild black holes, essentially, have
achieved heat death---one cannot extract energy from them without
perturbing them in an appropriate way.  Similarly, they cannot absorb
heat in s straightforward sense: if one absorbs ordinary heat from a
classical thermodynamical system, say, being thrown into it, then
after it settles down again to staticity it will once again have its
total mass equal to its irreducible mass (unless it acquires angular
momentum in the process, and so becomes a Kerr black hole).  In this
case, I think it still makes sense to say the black hole has absorbed
heat, in so far as, between the time the system is thrown in and the
time the black hole equilibrates again, its irreducible mass will not
be equal to its total mass.  The maximum of this difference, during
the equilibration process, will presumably equal the energy of the
system the black hole absorbed.  There are many challenges one could
reasonably pose to the approximations involved in attempting to carry
out such a calculation with anything approaching rigor (which I have
not done), but they are all the same sort of challenge one could pose
to the analogous problem in classical thermodynamics, so there is no
problem here peculiar to black-hole thermodynamics.

\subsection{Carnot-Geroch Cycles for Schwarzschild Black Holes}
\label{sec:schwarz-bh-carnot-cycles}

As I remarked at the end of \S\ref{sec:temp-ent-class-thermo}, the
strongest evidence that the formal equivalence of the laws of black
holes and those of ordinary thermodynamical systems in fact
constitutes a true physical equivalence, and that surface gravity is a
physical temperature and area a physical entropy, would consist in a
demonstration that black holes thermally couple with ordinary
thermodynamical systems in such a way that $\kappa$ plays the same
role in that coupling as ordinary temperature would if the system at
issue were coupling with another ordinary thermodynamical system and
not with a black hole, and the same for area.  My proposed
construction of the appropriate analogue for a Carnot cycle including
black holes, which I give in this subsection, will kill three birds
with one stone: not only will it show that $\kappa$ can be
characterized as the absolute temperature of the black hole using the
same arguments as classical thermodynamics uses to introduce the
absolute temperature scale; it will do so by showing that in the
coupling of black holes with ordinary thermodynamical systems,
$\kappa$ does in fact play the physical role of temperature and area
that of entropy; and it will have as a natural corollary the existence
of a universal constant that renders the proper physical dimensions to
surface gravity as a measure of temperature and area as a measure of
entropy.\footnote{I am grateful to Ted Jacobson for bringing to my
  attention after I wrote this paper the insightful analysis of
  \citeN{sciama-bhs-thermo}, in some ways quite similar to mine.  (See
  \citeNP{jacobson-intro-qft-cst} for a \emph{pr\`ecis} of Sciama's
  analysis.)  Sciama, however uses quantum systems all the way through
  and assumes that the analogy between black holes and ordinary
  thermodynamical systems is merely formal when one does not take
  quantum effects into account.}

I call the constructed process a ``Carnot-Geroch cycle'' both to mark
its difference from standard Carnot cycles, and because it relies
essentially on the mechanism at the heart of the most infamous example
in this entire field of study, Geroch's thought experiment of slowly
lowering towards a black hole a box filled with thermal matter, with
the argued consequence being that classical black holes must have
temperature absolute zero.\footnote{\label{fn:jiu-jitsu}According to
  Jakob Bekenstein (private correspondence) and Robert Wald
  (conversation), Geroch first proposed the example during a
  colloquium he gave at Princeton in 1970.  (Bekenstein tells me that
  he considers it the first attempt to attribute a temperature to a
  black hole.)  I cannot resist pointing out that my construction is
  essentially a jiu jitsu move against Geroch's original intent,
  turning the force of the example against itself, using Geroch's
  proposed mechanism to show that surface gravity really is a temp.}
(I discuss Geroch's original example and argue that it does not in
fact support the conclusions he wanted to draw from it in
\S\ref{sec:probs-poss-resol-insghts-qs} below.)  I will first sketch
the steps of the proposed cycle informally, then work through the
calculations.

\begin{description}
    \item[Reversible Carnot-Geroch Cycle Using a Schwarzschild Black
  Hole as a Heat Sink] \hspace*{1em}
  \begin{enumerate}
      \item start with a small, empty, essentially massless, perfectly
    insulating box ``at infinity'', one side of which is the outer
    face of a piston; in particular, the box is ``small'' in the sense
    that it will experience negligible tidal forces as it is lowered
    toward the black hole; very slowly (``quasi-statically'', so that
    the process is well approximated as an isentropic process) draw
    the piston back through the inside of the box, so filling the box
    with fluid from a large heat bath consisting of a large quantity
    of the fluid at fixed temperature $T_0$, so the fluid does work
    against the piston as it moves; when the piston has withdrawn part
    but not all of the way to the opposite side of the box, quickly
    seal the box, leaving the space opened by the piston filled with a
    mass of the fluid $M_0$ in thermal equilibrium at temperature
    $T_0$, and with entropy $S_0$; assume the entire energy of the box
    is negligible compared to the mass of the black hole
      \item very slowly, lower the box towards the black hole using an
    essentially massless rope; during this process, an observer inside
    the box would see nothing relevant change; in particular, as
    measured by an observer co-moving with the box, the temperature,
    volume and entropy of the fluid remain constant\footnote{The
      mass-density distribution of the fluid would change, increasing
      towards the side facing the black hole; this, however, does not
      affect the analysis, since this is what one expects for a system
      in thermal equilibrium in a quasi-static ``gravitational
      field''.  In any event, given our assumption about the size of
      the box, this effect would be negligible.}
      \item at a predetermined fixed proper radial distance from the
    black hole, stop lowering the box and hold it stationary
      \item very slowly, draw the piston back even further, so
    lowering the temperature of the fluid to a fixed, pre-determined
    value $T_1$ while keeping its entropy the same; the value of the
    temperature is to be fixed by the requirement that the change in
    total entropy vanishes during the next step (\emph{i}.\emph{e}.,
    entropy of black hole plus entropy of everything outside black
    hole does not change after the fluid is dumped into the black
    hole)
      \item open the box and eject the fluid out of it by using the
    piston to push it out, so the fluid falls into the black hole
    delivering positive mass-energy and positive entropy to it, and
    the piston returns to its initial state; by the way the
    temperature of the fluid was fixed in the previous step, this is
    an isentropic process \label{item:eject-fluid}
      \item pull the box back up to infinity (which takes no work, as
    the box now has zero mass-energy, and so zero weight), so it
    returns to its initial state
  \end{enumerate}
\end{description}
Because the total entropy remains constant during every step in the
process, these cycles are reversible in the sense of classical
thermodynamics.  Because the irreducible mass of the black hole
increases, however, it is not an irreversible process in the sense of
black-hole mechanics.\footnote{In \citeN{curiel-carnot-cyc-kerr-bhs},
  I propose another form of Carnot-Geroch cycle for a Kerr black hole,
  one that exploits its angular momentum in such a way as to make the
  process both reversible in the sense of classical thermodynamics and
  physically reversible according to black-hole mechanics.}

Now, let us make the following assumptions: first, that it makes sense
to attribute a physical temperature $T_\textsc{bh}$ and entropy
$S_\textsc{bh}$ to a black hole (though we do not yet know what they
are); second, that the entropy of ordinary thermodynamical systems and
the entropy of the black hole are jointly additive; and third, that
the appropriate temperature at which to eject the fluid into the black
hole for the entire cycle to be isentropic ($T_1$ in
step~\ref{item:eject-fluid}) is that one would expect for a thermally
equilibrated body in thermal contact with another at temperature
$T_\textsc{bh}$ sitting the given distance away in a nearly-static
gravitational field.  It will then follow that the physical
temperature must be $8 \pi \alpha \kappa$ and the physical entropy
$\displaystyle \frac{A}{\alpha}$, where $\kappa$ is the black hole's
surface gravity, $A$ its area, and $\alpha$ is a universal constant,
the analogue of Boltzmann's constant for black holes (to be derived
below).

Let the static Killing field in the spacetime be $\xi^a$ (timelike
outside the event horizon, null on it).  Let $\chi = (\xi^n
\xi_n)^{\half}$, and $a^a = (\xi^n \nabla_n \xi^a)/\chi^2$ be the
acceleration of an orbit of $\xi^a$.  Then a standard
calculation\footnote{See, \emph{e}.\emph{g}., \citeN{wald-gr}.}  shows
that
\[
\kappa = \lim(\chi a)
\]
where the limit is taken as one approaches the event horizon in the
radial direction, \emph{i}.\emph{e}., near the black hole $\chi a$ is
essentially the force that needs to be exerted ``at infinity'' to hold
an object so that it follows an orbit of $\chi^a$, which is to say, to
hold it so that it is locally stationary.  Thus $\chi$ is essentially
the ``redshift factor'' in a Schwarzschild spacetime.

Let the total energy content of the box when it is initially filled at
infinity be $E_0$ (as measured with respect to the static Killing
field).  In particular, $E_0$ includes contributions from the rest
mass of the fluid $M_0$, and from its temperature $T_0$ and entropy
$S_\text{b}$; let $W_0$ be the work done by the fluid as it pushes
against the piston in filling the box.  By the Gibbs relation and by
the First Law of thermodynamics, therefore, we can compute the
quantity of heat $Q_\text{b}$ initially in the box:
\[
Q_\text{b} = T_0 S_\text{b} = E_0 + W_0
\]
As the box is quasi-statically lowered to a proper distance $\ell$
from the event horizon, its energy as measured at infinity becomes
$\chi E_0$, where $\chi$ is the value of the redshift factor
at $\ell$.  Thus, the amount of work done at infinity in lowering the
box is
\[
W_\ell = (1 - \chi) E_0
\]
(Recall that we assumed the box to be so small that $\chi$ does not
differ appreciably from top to bottom.)  This is not standard
thermodynamical work, as the volume of the fluid, as measured by a
co-moving observer, has not changed.  It is rather work done by ``the
gravity of the black hole''.

Now, when the box is held at the proper distance $\ell$ from the black
hole and the piston slowly pushes or pulls so as to change the
temperature of the fluid from $T_0$ to $T_1$ (as measured locally),
the piston does work (as measured at infinity)
\[
W_1 = \chi (E_0 - E_1)
\]
where $E_1$ is the locally measured total energy of the fluid after
the fluid's (locally measured) volume has been changed by the piston.
When the fluid has reached the desired temperature $T_1$, the box is
opened and the piston pushes the fluid quasi-statically out of the
box, so it will fall into the black hole; in the process, the piston
does work $W_2$ (as measured at infinity).\footnote{One may worry that
  this process cannot be quasi-static, not even in principle, in so
  far as the phase-space volume available to the fluid as it is
  expelled from the box and before it is absorbed by the black hole
  is, in principle, unbounded, \emph{i}.\emph{e}., the entropy of the
  fluid increases by an arbitrary amount.  A superficial, but I think
  still adequate, answer to this problem is that one can arrange a
  telescopically extending mechanism from the box to the black hole to
  ensure that the volume available to the fluid never changes.  A
  deeper and I think more satisfying answer is that, when the fluid
  passes the event horizon, as all of it must do, its available
  phase-space volume only decreases, and arbitrarily so.  I thank Tim
  Maudlin for pushing me on this point.}  Now, by the First Law, the
total amount of energy the fluid has as it leaves the box is
\begin{equation}
  \label{eq:energy-leaves-box}
  E_1 - \frac{W_2}{\chi} = T_1 S_\text{b}
\end{equation}
as measured locally.  

In order to compute the total amount of energy and the total amount of
heat dumped into the black hole as measured at infinity, we must
compute the temperature of the box as measured from there.  It is a
standard result \cite[p.~318]{tolman-rel-thermo-cosmo} that the
condition for a body at locally measured temperature $T$ to be in
thermal equilibrium in a strong, nearly static gravitational field is
that the temperature measured ``at infinity'' be $\chi T$.  Thus the
temperature of the box as measured from infinity will be $\chi T_1$.
It follows from equation~\eqref{eq:energy-leaves-box}, therefore, that
the total amount of heat dumped into the black hole is
\[
\chi T_1 S_\text{b} = \chi E_1 - W_2
\]
But $\chi E_1 = \chi E_0 - W_1$ and $\chi E_0 = E_0 -
W_\ell$, so
\[
\chi T_1 S_\text{b} = E_0 - W_\ell - W_1 - W_2
\]
The expression on the righthand side of the last equation, however, is
just the total amount of energy in the box as measured at infinity,
and so $\chi T_1 S_\text{b}$ is the total amount of energy the black
hole absorbs, as measured from infinity, which is entirely in the form
of heat.

Now, because we have assumed that the entropy for the fluid and for
the black hole is additive, the total change in entropy is
\[
\Delta S = -S_\text{b} + \frac{\chi T_1 S_\text{b}}{T_\textsc{bh}}
\]
For the process to be isentropic, 
\[
\Delta S = 0
\]
and so
\begin{equation}
  \label{eq:isentropic-cond}
  \frac{\chi T_1 S_\text{b}}{T_\textsc{bh}} = S_\text{b}
\end{equation}
Thus, $\displaystyle T_1 = \frac{T_\textsc{bh}} {\chi}$, precisely the
temperature one would expect for a thermally equilibrated body in
thermal contact with another body at temperature $T_\textsc{bh}$ a
redshift distance $\chi$ away.  Write $Q_\textsc{bh}$ for the amount
of heat the black hole absorbs ($= \chi T_1 S_\text{b}$), so
equation~\eqref{eq:isentropic-cond} becomes
\[
\frac{Q_\textsc{bh}}{T_\textsc{bh}} = S_\text{b}
\]

Now, in the limit as the box, and so the heat and entropy it contains,
becomes very small (while the temperature remains constant), we may
think of this as an equation of differentials,
\begin{equation}
  \label{eq:temp-integ-fact}
  \frac{\text{d} Q_\textsc{bh}}{T_\textsc{bh}} = \text{d} S_\text{b}
\end{equation}
This expresses the well known fact that temperature plays the role of
an integrating factor for heat.  Since $\text{d} Q_\textsc{bh}$ is the
change in mass of the black hole, $\text{d} M_\textsc{bh}$, due to its
being the entirety of the energy absorbed, there follows from the
First Law of black-hole mechanics\footnote{At least two conceptually
  distinct formulations of the First Law of black-hole mechanics
  appear in the literature, what (following \citeNP[ch.~6,
  \S2]{wald-qft-cst}) I will call the physical-process version and the
  equilibrium version.  The former fixes the relations among the
  changes in an initially stationary black hole's mass, surface
  gravity, area, angular velocity, angular momentum, electric
  potential and electric charge when the black hole is perturbed by
  throwing in an ``infinitesimally small'' bit of matter, after the
  black hole settles back down to stationarity.  The latter considers
  the relation among all those quantities for two black holes in
  ``infinitesimally close'' stationary states, or, more precisely, for
  two ``infinitesimally close'' black-hole spacetimes.  Clearly, I am
  relying on the physical-process version, for the most thorough and
  physically sound discussion and proof of which see
  \citeN{wald-gao-proc-1st-genl-2nd-charged-rot-bhs}.}
\begin{equation}
  \label{eq:surfg-integ-fact}
  \frac{8\pi \/ \text{d} Q_\textsc{bh}}{\kappa} = \text{d} A
\end{equation}
Thus, $\kappa$ is also an integrating factor for heat.  It is a well
known theorem that if two quantities are both integrating factors of
the same third quantity, the ratio of the two must be a function of
the quantity in the total differential, and so in this case
\begin{equation}
  \label{eq:ratio-k-T}
  \frac{T_\textsc{bh}} {\kappa} = \psi(A)
\end{equation}
for some $\psi$.  (It is also the case that $\displaystyle
\frac{T_\textsc{bh}} {\kappa} = \phi(S_\text{b})$ for some $\phi$, but
we will not need to use that.)  It follows from
equations~\eqref{eq:temp-integ-fact} and \eqref{eq:surfg-integ-fact}
that
\begin{equation}
  \label{eq:A-func-S}
  \frac{1}{8\pi} \psi(A) \text{d} A = \text{d} S_\text{b}
\end{equation}
and so integrating this equation yields the change in the black hole's
area, $\Delta A$ as a function of $S_\text{b}$, say $\Delta A = \theta
(S_\text{b})$.  (From hereon, we fix some arbitrary standard value for
$A$, and so drop the `$\Delta$'.)

In order to complete the argument, and make explicit the relation
between $A$ and $S_\text{b}$, and at the same time fix the relation
between $\kappa$ and $T_\textsc{bh}$, consider two black holes very
far apart, and otherwise isolated, so there is essentially no
interaction between them.  Perform the Geroch-Carnot cycle on each
separately.  Let $A_1$ and $A_2$ be their respective areas, $\theta_1$
and $\theta_2$ the respective functions for those areas expressed
using $S_\text{b1}$ and $S_\text{b2}$, the respective entropies dumped
into the black holes by the cycles, and let $\theta_{12}
(S_\text{b12})$ be the function for the total area of the black holes
considered as a single system, expressed using the total entropy
$S_\text{b12}$ dumped into the system.  Both the total area of the
black holes and the total entropy dumped in are additive (since the
black holes, and so the elements of the Carnot-Geroch cycles, have
negligible interaction), \emph{i}.\emph{e}.,
\[
\theta_1 (S_\text{b1}) + \theta_2 (S_\text{b2}) = \theta_{12}
(S_\text{b12}) = \theta_{12} (S_\text{b1} + S_\text{b2})
\]
Differentiate each side, first with respect to $S_\text{b1}$ and then
with respect to $S_\text{b2}$; because $\theta_{12}$ is symmetric in
$S_\text{b1}$ and $S_\text{b2}$,
\[
\frac{\text{d} \theta_1}{\text{d} S_\text{b1}} = \frac{\text{d}
  \theta_2}{\text{d} S_\text{b2}}
\]
Since the parameters of the two black holes and the two cycles are
arbitrary, it follows that there is a universal constant $\alpha$ such
that
\[
\frac{\text{d} \theta}{\text{d} S_b} = \frac{\text{d} A}{\text{d} S_b}
= \alpha
\]
for all Schwarzschild black holes.  It now follows directly from
equations~\eqref{eq:ratio-k-T} and \eqref{eq:A-func-S} that
\begin{equation}
  \label{eq:bh-temp-def}
  T_\textsc{bh} = 8 \pi \alpha \kappa
\end{equation}
and from equation~\eqref{eq:isentropic-cond} that
\begin{equation}
  \label{eq:bh-ent-def}
  S_\textsc{bh} = \frac{A}{\alpha}
\end{equation}
up to an additive constant we may as well set equal to
zero.\footnote{\label{fn:natl-unit-bh-ent}In contradistinction to
  classical thermodynamical systems, geometrized units for the entropy
  of black holes can be naturally constructed: let a natural unit for
  mass be, say, that of a proton; then one unit of entropy is that of
  a Schwarzschild black hole of unit mass.  Why does classical
  black-hole thermodynamics allow for the construction of a natural
  unit for entropy when purely classical, non-gravitational
  thermodynamics does not?}  $\alpha$ is guaranteed by construction to
have the proper dimensions to give $T_\textsc{bh}$ the physical
dimension of temperature (mass, in geometrized units), and
$S_\textsc{bh}$ the physical dimension of entropy (dimensionless, in
geometrized units).

As a consistency check, it is easy to compute that the total work
performed in the process,
\[
W_T = W_0 + W_\ell + W_1 + W_2
\]
equals the total change in heat of the box during the process,
$Q_\text{b} - \chi T_1 S_\text{b}$, exactly as one should expect for a
Carnot cycle.  One can use the total work, then, to define the
efficiency of the process in the standard way,
\[
\eta \coloneq \frac{W_T} {Q_\text{b}} = 1 - \frac{\chi T_1 S_\text{b}}
{Q_\text{b}}
\]
from which it follows that
\[
\eta = 1 - \frac{8 \pi \alpha \kappa}{T_0}
\]
Thus, one can use the standard procedure for defining an absolute
temperature scale based on the efficiency of Carnot cycles, and one
concludes that the absolute temperature of the black hole is indeed $8
\pi \alpha \kappa$.

Unfortunately, one cannot use similar arguments as in the classical
case to prove the analogue of theorem~\ref{thm:revers-eng-eff}, as the
Carnot-Geroch Cycle for Schwarzschild black holes is not reversible in
the physical sense.  Under restricted conditions, however, the
Carnot-Geroch cycle for Kerr black holes \emph{is} physically
reversible, and so in that case one can use the classical arguments to
prove the analogue of theorem~\ref{thm:revers-eng-eff}, as I plan to
discuss in future work \cite{curiel-carnot-cyc-kerr-bhs}.

\subsection{The Generalized Clausius and Kelvin Postulates for Black
  Holes}
\label{sec:bh-clausius-kelvin}

Although I consider the construction of the Carnot-Geroch Cycle and
the arguments based on it to be the most decisive in favor of
conceiving of classical black holes as truly thermodynamical objects,
I think it is still worthwhile to show that the appropriately
translated analogues of the Clausius and Kelvin Postulates hold for
black holes as well.  Because those Postulates provide the ground for
all ways of introducing temperature and entropy in classical
thermodynamics, to show that they hold of black holes as well will
show that the physical behavior of black holes conforms as closely as
possible to that of classical thermodynamical in all fundamental
respects.

The standard arguments in favor of the Clausius and Kelvin postulates
(as given, \emph{e}.\emph{g}., in \citeNP[ch.~3]{fermi-thermo}), which
rely on the impossibility of constructing a \emph{perpetuum mobile} of
the second kind, do not translate straightforwardly into the context
of general relativity, where there is no general principle of the
conservation of energy.  Remarkably enough, however, one can still
give arguments for them at least as strong as those given for their
analogues in classical thermodynamics.
\begin{postulate}[Generalized Clausius Postulate for Black Holes]
  \label{post:bh-clausius}
  For any two systems, at least one of which is a stationary black
  hole, a transformation whose only final result is that a ``quantity
  of heat'' (as defined in \S\ref{sec:irred-mass-free-energy-heat}) is
  transferred from the system with lower temperature (surface gravity)
  to the one of higher temperature (surface gravity) is impossible.
\end{postulate}
Assume that initially the black hole is at the lower temperature, and
that such a transformation as described in the antecedent of the
theorem were possible.  Then the change in irreducible mass of the
black hole would have to be strictly greater than the change in its
total mass during the interaction, with no other change in the
spacetime than that another system absorbed heat.  In particular, its
irreducible mass must increase.  However, it follows from equation
\eqref{eq:area-irr-mass} that an increase in irreducible mass must
yield an increase in the black hole's area, and so its entropy,
violating the assumption that nothing else thermodynamically relevant
in the spacetime changed.  Analogous reasoning in the case where the
black hole is initially at a higher temperature shows that the
irreducible mass would also have to change in such a process.

\begin{postulate}[Generalized Kelvin Postulate for Black Holes]
  \label{post:bh-kelvin}
  A transformation whose only final result is that a ``quantity of
  heat'' (as defined in \S\ref{sec:irred-mass-free-energy-heat}) is
  extracted from a stationary black hole and transformed entirely into
  work is impossible.
\end{postulate}
The argument is essentially the same as for the Clausius Postulate for
black hole.  Again, for such a process to occur, the irreducible mass
of the black hole would have to increase, but that would necessitate a
change in the area of the black hole, violating the conditions of the
theorem.

\section{Problems, Possible Resolutions, Possible Insights, and
  Questions}
\label{sec:probs-poss-resol-insghts-qs}

I conclude the paper with a brief discussion of some \emph{prima
  facie} problems with my arguments, suggestions for their
resolutions, an examination of what insights my conclusions, if
correct, may offer, and some general questions that I think need to be
addressed, possibly with the help of my arguments and conclusions.

An obvious complaint against the argument based on the construction of
the Carnot-Geroch Cycle is that it is circular: why assume a classical
black hole has an entropy in the first place?  The best answer to this
is implicit in the questions Wheeler initially posed in the late 1960s
that inspired the entire field of black-hole thermodynamics in the
first place: if we don't assume black holes have entropy, then we
would, with effortless virtuosity, be able to achieve arbitrarily
large violations of the Second Law of thermodynamics.  The world
external to a black hole is isolated from the interior of the black
hole.  So, take your favorite highly entropic system and throw it into
a black hole: the entropy of that system vanishes from the external
world, so lowering the total entropy of an isolated system.  The only
escape from this possibility is to assign the black hole itself an
entropy in such a way that, when an ordinary entropic system passes
into a black hole, then the black hole's entropy increases at least as
much as the entropy of the system entering it.  This postulate is
generally referred to as the Generalized Second Law: the total entropy
of the world, \emph{viz}., the entropy of everything outside black
holes plus the entropy of black holes, never decreases
\cite{bekenstein-bh-ent,bekenstein-genl-2nd-law-bh-phys}.

This attempt to answer the first problem leads naturally to the next,
possibly the most serious potential problem: the derivation of the
relation between black-hole entropy and area based on the
Carnot-Geroch cycle does not by itself guarantee that there is no
process that violates the Generalized Second Law.  In particular,
though in footnote~\ref{fn:jiu-jitsu} I claimed to turn Geroch's
infamous thought-experiment on its head, nothing seems to preclude
Geroch's original use of it to argue that, were classical black holes
to have physical temperature, it would have to be absolute zero
independently of what value its surface gravity had.  If one arranges
matters just so, the weight lifted by the lowering of the box will
have extracted \emph{all} the energy content of the box when it
reaches the event horizon; one can then dump into the black hole the
stuff in the box, which still has its original entropy but zero
mass-energy; thus, one will have converted thermal energy into work
with 100\% efficiency, implying the black hole must have temperature
absolute zero.  Because the matter dumped into the black hole has no
mass-energy, the area of the black hole does not increase; because the
matter still has its original entropy, however, the total entropy of
the world outside the event horizon has decreased, thus violating the
Generalized Second Law.

There are (at least) two possible responses.\footnote{Perhaps the most
  influential response in the physics literature to this problem is
  given by \citeN{unruh-wald-acc-rad-gsl}.  I will not consider their
  response, as it inextricably relies on quantum effects.}  First, one
can note that the procedure requires measurements of arbitrarily fine
precision: the violation of the Generalized Second Law occurs
\emph{only} if the matter has \emph{exactly} zero stress-energy when
it is released \emph{precisely} when the box is contiguous with the
event horizon.  Otherwise, the area of the black hole will increase,
and will always do so in way so as to preserve the Generalized Second
Law.  If one holds that classical thermodynamics is only an effective
theory in the first place, as seems reasonable, then the notion of
arbitrarily precise measurements never gets off the
ground.\footnote{This is not the same issue as arises with arguments
  over the possibility of a Maxwell demon, though the demon may have
  to make arbitrarily fine measurements in order to function.  The
  Maxwell demon in classical thermodynamics will eventually
  thermalize, and so one will have to continually produce a new demon
  from a low-entropy source in order to produce arbitrarily large
  deviations from the Second Law, whereas nothing in the black-hole
  case thermalizes, so if one could make arbitrarily precise
  measurements then one should be able to systematically produce
  arbitrarily large deviations from the Generalized Second Law.}

The second possible response accepts the possibility of arbitrarily
precise measurements in the context of classical thermodynamics.  If
one allows the possibility of such measurements, however, then it is
not a justified idealization to ignore the stress-energy contained in
the rope holding the box above the black hole.  One may be justified
in treating the rope as having \emph{initially} zero stress-energy, as
an idealization, but once the box approaches the black hole, the
internal tension in the rope will become a non-trivial momentum flux
(as different parts of the rope, at different distances from the
horizon, pull on each other with different force), and so one has to
take account of that stress-energy.\footnote{See
  \citeN{thorne-etal-bhs-membrane} for detailed calculations taking
  account of the rope's stress-energy during such a lowering process.}
One will, therefore, never be able to get the internal energy of the
box exactly to zero before one dumps the entropic stuff in it into the
black hole.

There is, however, another possible mechanism for producing
arbitrarily large violations of the Generalized Second Law if one
treats classical black holes as truly thermodynamical objects.  Put a
Kerr black hole in a reflecting box and pervade the box with thermal
electromagnetic radiation at a lower (Planck) temperature than the
classical Bekenstein-Hawking temperature of the black hole.  The black
hole will eventually absorb the thermal radiation: heat would
spontaneously transfer from a system at a lower temperature to one at
a higher temperature, a seeming violation of the Generalized Second
Law.\footnote{I thank Robert Wald for proposing this example to me.}
First, one should note that this is \emph{not} a violation of the
Generalized Clausius Postulate, as the irreducible mass and so the
area of the black hole increase after absorption.  If one takes the
Generalized Clausius Postulate as the appropriate formulation of the
Generalized Second Law in the context of classical black-hole
mechanics and thermodynamics, as the ordinary Clausius Postulate is in
classical thermodynamics alone, then there is no violation of the
Generalized Second Law.

Although I think this response is correct, it may still seem
unsatisfying in so far as it still looks as though there may be
violations of the generalized principle of entropy increase
(\emph{i}.\emph{e}., what is standardly called the Generalized Second
Law in the literature).  Now, in order to justify the claim that this
constitutes a violation of the generalized principle of entropy
increase, in the sense that the sum of the external entropy and
black-hole area is less after absorption than it was before, one has
to verify that in fact the increase in the black hole's entropy after
it absorbs all the radiation (and settles back down to equilibrium)
will be less than the entropy originally contained in the radiation.
The energy content of the radiation is proportional to its (Planck)
temperature raised to the fourth power, $T_P^4$, and its entropy to
$T_P^3$.  For simplicitly, assume that those powers of $T_P$ just are
the radiation's temperature and entropy respectively.  The increase in
the entropy of the black hole then is its increase in area.  Because
$A = E^2$ (ignoring constant factors), $\Delta A = (E + T_P^4)^2 - E^2
= T_P^8 - 2ET_P^4$, which may be greater or less than $T_P^3$
depending on the values of $E$ and $T_p$ (and the ignored
constants).\footnote{This simple calculation ignores the fact that the
  black hole will emit gravitational radiation as it is perturbed by
  the in-falling electromagnetic radiation.  It is possible that,
  though the resulting gravitational radiation will carry very little,
  essentially negligible, energy, it may still contain non-negligible
  entropy, if purely gravitational entropy in such a form has the same
  outlandishly high relative values as it does for black holes, as the
  arguments of \citeN{clifton-et-gravl-ent-propos} suggest.  If that
  is correct, then it may be that, when the entropy of the
  gravitational radiation is taken into account, it would not be
  possible to violate even the principle of entropy increase by this
  mechanism.}

Now, one may want to say that this shows that classical black holes
cannot be conceived consistently as thermodynamical objects, in so far
as we may have here a case of heat spontaneously flowing from a colder
to a hotter system, irrespective of whether or how total entropy
changes.  In the event, however, the violation turns out to depend
crucially on the fact that in this case one models the radiation as a
\emph{quantum} system, while treating the black hole as a purely
classical system---for one will get \emph{exactly the same behavior}
for any classical thermodynamical system put in place of the classical
black hole.  Put an ordinary classical thermodynamical system
(\emph{e}.\emph{g}., a classical fluid) in a reflecting box and
pervade the box with thermal radiation at a lower (Planck)
temperature, $T_P$, than that of the classical fluid, $T$.  Because
the radiation is modeled using quantum mechanics and the fluid using
classical thermodynamics, it is ambiguous how to model their
interaction and joint evolution.  There are two possibilities.  First,
one may assume that the fluid will absorb the radiation.  Second, one
may assume it does not.  In either case, because the fluid is modeled
using classical thermodynamics, it will not \emph{emit} any radiation,
black body or otherwise.  It is only the first possibility that
interests us here, because that is the case analogous to the black
hole's behavior.

To determine whether or not there is a violation of the generalized
principle of entropy increase, note first that the change in the
fluid's temperature, after it absorbs the radiation, will be $C_V{}
T_P^4$, where $C_V$ is the fluid's specific heat.  Because the fluid's
initial entropy is $E/T$ (from the Gibbs relation, $E$ being the
inital total energy of the system), the change in the fluid's entropy
will be
\[
\frac{E + T_P^4} {T + C_V{} T_P^4} - \frac{E}{T}
\]
This may be greater than or less than $T_P^3$ depending on the values
of $E$, $T$, $C_V$ and $T_P$.  (This may be easily seen from the fact
that the first term in the sum diverges as $T_p$ goes to zero, whereas
$T_p^3$ does not, and the second term diverges as $T_p$ goes to
infinity more slowly than does $T_p^3$ while the first term approaches
a constant.)  This, however, is exactly the situation as for the
classical black hole.  Thus, in fact, the classical black hole behaves
\emph{exactly like a classical thermodynamical system}, which is the
only conclusion I am arguing for.

One may want to conclude that this is not a satisfactory result:
\emph{any} systematic way to violate the generalized principle of
entropy increase should be considered illegitimate, irrespective of
the constraints one has to impose on the systems one models, and how
one models them, to arrange it.\footnote{I believe this is Wald's
  reaction to the situation.}  I think, however, that these arguments
rather show that it is simply inconsistent, in the context of
thermodynamics, to model one system using quantum mechanics and
another using classical thermodynamics, when one treats the systems as
interacting, as I suggested in \S\ref{sec:standard-argument}.  Another
possible lesson one may want to draw from these arguments, one I am
sympathetic to, is that the appropriate form of the Generalized Second
Law should not be the generalized principle of entropy increase, but
rather the Generalized Clausius Postulate.

Before turning to what we may learn from my conclusions, if they are
correct, I consider a few more possible problems, none of which I
consider severe.  Indeed, the resolution of all of them lies in
showing that, again, the proposed problem really is a problem for
treating classical black holes as thermodynamical systems if and only
if it is also a problem for ordinary classical thermodynamical
systems---again, classical black holes behave in every way like
ordinary thermodynamical systems.

As is well known, the surface gravity $\kappa$ is well defined only
for stationary black holes; does this mean that my analysis cannot
apply to non-stationary black holes?  Yes, it is the case that my
analysis cannot apply to non-stationary black holes, but that is no
problem.  Non-stationary black holes are ones out of equilibrium, and
so this presents the same situation as obtains in classical
equilibrium thermodynamics.  I think we often forget that, strictly
speaking, temperature in ordinary thermodynamics is well defined only
for bodies in (or quite close to) thermal equilibrium.  One way to see
this is to note that, for systems far from equilibrium, different
kinds of thermometric device will return very different readings, as
fine details of their different couplings to the system which are
negligible for equilibrium systems become non-trivial, in particular
due to phenomena manifesting themselves at temporal and spatial scales
below the hydrodynamic scale.\footnote{See, \emph{e}.\emph{g}.,
  \citeN[\S\S4.1--4.4, pp.~24--9]{benedict69}.  This reference is not
  the most up-to-date with regard to the international agreement on
  defining the standard, practical methods for the determination of
  temperature, but I have found no better reference for the nuts and
  bolts of thermometry.  See \citeN[\S3.4]{curiel-theoryexp-ivf-pde}
  for a discussion of the details.}

Another problem is that it seems as though we can attribute heat to a
Schwarzschild black hole only when it is being perturbed.  Again, the
situation is in fact much the same as in classical thermodynamics,
wherein it never makes sense to attribute a definite quantity of heat
to an isolated system in equilibrium.  The only definite claims we can
make, as Maxwell himself so insightfully and eloquently discussed
(footnote~\ref{fn:max-heat}), are about the quantification of heat
\emph{transfer}.  In any event, one \emph{can} extract both ``heat''
and work from a Schwarzschild black hole by perturbing it; indeed,
this is in excellent analogy with ordinary thermodynamical systems
that have reached heat death, from which heat and work can be
extracted only if one perturbs them properly.  In fact, the analogy is
even better than that brief remark suggests: stationary classical
black holes do not ``radiate heat'', but neither do ordinary classical
thermodynamical systems in equilibrium; classical systems exchange
heat only when they are in direct contact (contiguous) with another
system at a different temperature, but the same holds for stationary
classical black holes, in so far as their immediately contiguous
environment is ``at the same temperature'', \emph{viz}., has
essentially the same effective surface gravity as measured at
infinity, as the black hole does.  Still, one may protest, in the
construction of the Carnot-Geroch Cycle, I ignored perturbations to
the black hole from the lowering of the box, so how can one say, given
my definitions and arguments, that energy was extracted from it?
Given the assumption that the total energy of the box is negligible
compared to the mass of the black hole, I claim it is a good
approximation to ignore any perturbations to the black hole while
still accounting for the (relatively negligible) amount of energy the
box gains by being lowered through the black hole's ``gravitational
field''.

Another potential problem: it is clear that black holes have, by the
standard definition, negative specific heat, since their surface
gravity decreases as their mass-energy increases.  Standard arguments,
however, conclude that two bodies with negative specific heat cannot
thermally equilibrate.  There is, though, a hidden assumption in the
standard arguments, to wit, ``conservation of heat''---it is always
assumed, that is to say, that for two bodies in thermal contact one
can gain heat only if the other loses it, and that in the same amount.
Heat, however, is not a substance, as everyone from
\citeN{maxwell-theory-heat-1888} to \citeN{planck97/26} to
\citeN{sommerfeld-thermo} is at pains to emphasize, and so obeys no
conservation law.  There is no reason why two bodies with different
temperatures in thermal contact cannot both ``gain or lose heat from
or to each other'' at the same time.  When two black holes in
quasi-stationary orbit\footnote{There are no solutions to the Einstein
  field equation representing two Kerr black holes in stable orbit
  about each other \cite{manko-ruiz-exact-soln-dbl-kerr-equil}.}
about each other equilibrate, the temperatures of both bodies
simultaneously decrease as they both gain heat from the other, the one
of higher temperature decreasing more quickly than the other, so they
will eventually reach the same temperature.

A potentially more serious problem with my analysis is that it is
difficult to see what sense can be made of ``exchange'' between a
global energetic quantity (in the case of stationary, asymptotically
flat black holes, ADM mass) on the one hand, and localized
stress-energy of ordinary systems on the other.  A more poignant way
of posing the problem is to note that gravitational energy is strictly
non-local in the precise sense that there is no such thing as a
gravitational stress-energy tensor
\cite{curiel-geom-objs-nonexist-sab-uniq-efe}, and so it satisfies no
general conservation law.  How, then, can one talk about exchange for
such a \emph{recherch\'e} quantity?\footnote{I thank Jim Weatherall
  for pushing me on this point.}  There are, I think, two responses to
this problem, one stronger than the other.  The first, weaker,
response is that one always has in place a quasi-local notion of
mass-energy in stationary and axisymmetric spacetimes, which suffices
for the purposes of my arguments, just as it does in Newtonian
gravitational theory (\emph{\`a la} the ``Poynting integral'' of
\citeNP{bondi-phys-char-grav-wvs}).  The stronger response, which is
more to the point, is that neither is heat a localized form of energy
in classical thermodynamics---it is not a perfect differential (as the
discussion of \citeNP{sommerfeld-thermo} makes particularly clear),
and so it also has no corresponding conservation law---just like
gravitational energy---and yet we feel no inconsistency in talking
there about exchange of energy for a quantity that can be represented
only as a total magnitude, with no corresponding localized density.
Sauce for the goose is surely sauce for the gander.

My arguments, I think, have not only residual possible problems; they
also open the possibility for real insight into existing questions
about black-hole mechanics and thermodynamics.  Although the following
is not a problem peculiar to my analysis, it is a general one in the
field I believe my analysis can give some insight into.  Black holes
have enormous entropy, far more than any reasonably conceivable
material system that could form them on collapse
\cite{penrose-sing-time-asym}.  There must, therefore, be a
correspondingly enormous and discontinuous jump in entropy when a
collapsing body passes the point at which an event horizon forms.  How
can one explain that?  It is here that I believe my old-fashioned
approach to entropy bears some of its sweetest fruit.  More modern
characterizations of entropy, whether of a Boltzmannian, Gibbsian,
von-Neumann-like, or Shannon-like form, have no explanation for this
jump.  If, however, one conceives of entropy as a measure of how much
work it takes to extract energy from a system, how much free energy a
system has, what forms its internal energy (as opposed to free energy)
are in, then black holes have enormous energy, only a very small
amount of which is extractible, and there is a clear physical
discontinuity in extractability of energy when an event horizon forms.

I leave the reader with a question concerning this entire field that,
though not peculiar to my arguments here, I feel strongly needs to be
investigated further by both philosophers and physicists.  The Laws of
thermodynamics are empirical generalizations, indeed, the paradigm of
such.  I know of no other fundamental propositions in physics whose
support comes \emph{entirely} from experimental evidence, with not
even the suggestion of the possibility of a formal derivation from
``deeper'' physical principles.  Also, I know of no other
propositions, with the possible exception of the Newtonian
inverse-squared distance dependence of gravitational attraction
between two bits of matter, that are more deeply entrenched
empirically than the Laws of thermodynamics.  But, entirely to the
contrary, and with the exception only of the Third Law (which is also
the most weakly supported by experimental evidence in classical
thermodynamics), all the Laws of black-hole mechanics are theorems of
differential geometry.  They require no input from physical theory at
all.  One will sometimes see the claim that one or the other of the
Laws requires the assumption of the Einstein field equation, but this
is not true: all the Laws are independent of the Einstein field
equation in the strong sense that one can assume its negation and
still derive the Laws; the Einstein field equation enters only when
one wants to give a physical interpretation of the quantities involved
by way of its asserted relation between the Ricci tensor and the
stress-energy tensor of matter.\footnote{See
  \citeN{curiel-primer-econds} for a thorough discussion.}  So how can
laws that, in one context, are nothing but empirical generalizations,
magically transform into mathematical theorems when extended into a
new context?


\begin{thebibliography}{}

\bibitem[\protect\citeauthoryear{Aizenman and Lieb}{Aizenman and
  Lieb}{1981}]{aizenman-lieb-3rd-law-degen-grnd-state}
Aizenman, M. and E.~Lieb (1981).
\newblock The {T}hird {L}aw of thermodynamics and the degeneracy of the ground
  state for lattice systems.
\newblock {\em Journal of Statistical Physics\/}~{\em 24\/}(1), 279--297.
\newblock \href{http://dx.doi.org/10.1007/BF01007649} {doi:10.1007/BF01007649}.

\bibitem[\protect\citeauthoryear{Bardeen, Carter, and Hawking}{Bardeen
  et~al.}{1973}]{bardeen-carter-hawking73}
Bardeen, J., B.~Carter, and S.~Hawking (1973).
\newblock The four laws of black hole mechanics.
\newblock {\em Communications in Mathematical Physics\/}~{\em 31\/}(2),
  161--170.
\newblock \href{http://dx.doi.org/10.1007/BF01645742} {doi:10.1007/BF01645742}.

\bibitem[\protect\citeauthoryear{Bekenstein}{Bekenstein}{1973}]{bekenstein-bh-ent}
Bekenstein, J. (1973).
\newblock Black holes and entropy.
\newblock {\em Physical Review D\/}~{\em 7}, 2333--2346.
\newblock \href{http://dx.doi.org/10.1103/PhysRevD.7.2333}
  {doi:10.1103/PhysRevD.7.2333}.

\bibitem[\protect\citeauthoryear{Bekenstein}{Bekenstein}{1974}]{bekenstein-genl-2nd-law-bh-phys}
Bekenstein, J. (1974).
\newblock Generalized {S}econd {L}aw of {T}hermodynamics in black-hole physics.
\newblock {\em Physical Review D\/}~{\em 9}, 3292--3300.
\newblock \href{http://dx.doi.org/10.1103/PhysRevD.9.3292}
  {doi:10.1103/PhysRevD.9.3292}.

\bibitem[\protect\citeauthoryear{Benedict}{Benedict}{1969}]{benedict69}
Benedict, R. (1969).
\newblock {\em Fundamentals of Temperature, Pressure and Flow Measurements}.
\newblock New York: John Wiley \& Sons, Inc.

\bibitem[\protect\citeauthoryear{Bondi}{Bondi}{1962}]{bondi-phys-char-grav-wvs}
Bondi, H. (1962).
\newblock On the physical characteristics of gravitational waves.
\newblock In A.~Lichnerowicz and A.~Tonnelat (Eds.), {\em Les Th\'eories
  Relativistes de la Gravitation}, Number~91 in Colloques Internationaux, pp.\
  129--135. Paris: Centre National de la Recherche Scientifique.
\newblock Proceedings of a conference held at Royaumont in June, 1959.

\bibitem[\protect\citeauthoryear{Bredberg, Keeler, Lysov, and
  Strominger}{Bredberg et~al.}{2011}]{bredberg-etal-navst-einst}
Bredberg, I., C.~Keeler, V.~Lysov, and A.~Strominger (2011).
\newblock From {N}avier-{S}tokes to {E}instein.
\newblock \href{http://arxiv.org/absolute/hep-th/1101.2451}
  {arXiv:hep-th/1101.2451v2}.

\bibitem[\protect\citeauthoryear{Brown and Uffink}{Brown and
  Uffink}{2001}]{brown-uffink-minus-first}
Brown, H. and J.~Uffink (2001, December).
\newblock The origins of time-asymmetry in thermodynamics: {T}he minus first
  law.
\newblock {\em Studies in History and Philosophy of Modern Physics\/}~{\em
  32\/}(4), 525--538.
\newblock
  \href{http://dx.doi.org/10.1016/S1355-2198(01)00021-1}{doi:10.1016/S1355-2198(01)00021-1}.

\bibitem[\protect\citeauthoryear{Carath\'eodory}{Carath\'eodory}{1909}]{caratheodory-grund-thermo}
Carath\'eodory, C. (1909).
\newblock Untersuchungen \"uber die {G}rundlagen der {T}hermodynamik.
\newblock {\em Mathematische Annalen\/}~{\em 67\/}(3), 355--386.
\newblock \href{http://dx.doi.org/10.1007/BF01450409} {doi:10.1007/BF01450409}.

\bibitem[\protect\citeauthoryear{Carter}{Carter}{1973}]{carter73}
Carter, B. (1973).
\newblock Black hole equilibrium states.
\newblock In B.~DeWitt and C.~DeWitt (Eds.), {\em Black Holes}, pp.\  56--214.
  New York: Gordon and Breach.

\bibitem[\protect\citeauthoryear{Christodoulou}{Christodoulou}{1970}]{christodoulou70}
Christodoulou, D. (1970).
\newblock Reversible and irreversible transformations in general relativity.
\newblock {\em Physical Review Letters\/}~{\em 25\/}(22), 1596--1597.
\newblock \href{http://dx.doi.org/10.1103/PhysRevLett.25.1596}
  {doi:10.1103/PhysRevLett.25.1596}.

\bibitem[\protect\citeauthoryear{Clifton, Ellis, and Tavakol}{Clifton
  et~al.}{2013}]{clifton-et-gravl-ent-propos}
Clifton, T., G.~Ellis, and R.~Tavakol (2013).
\newblock A gravitational entropy proposal.
\newblock {\em Classical and Quantum Gravity\/}~{\em 30\/}(12), 125009.
\newblock \href{http://dx.doi.org/10.1088/0264-9381/30/12/125009}
  {doi:10.1088/0264-9381/30/12/125009}. Preprint available at
  \href{http://arxiv.org/abs/1303.5612} {arXiv:1303.5612v2 [gr-qc]}.

\bibitem[\protect\citeauthoryear{Curiel}{Curiel}{2010}]{curiel-theoryexp-ivf-pde}
Curiel, E. (2010).
\newblock On the formal consistency of theory and experiment, with applications
  to problems in the initial-value formulation of the partial-differential
  equations of mathematical physics.
\newblock This paper is a corrected and clarified version of the third chapter
  of the author's Ph.D. dissertation, \emph{Three Papers on How Physics Bears
  on Philosophy, and How Philosophy Bears on Physics} (Philosophy Department,
  University of Chicago, 2005). An updated version is available online at
  \url{http://strangebeautiful.com/papers/curiel-theory-experiment.pdf}.

\bibitem[\protect\citeauthoryear{Curiel}{Curiel}{2014a}]{curiel-carnot-cyc-kerr-bhs}
Curiel, E. (2014a).
\newblock {C}arnot-{G}eroch cycles for {K}err black holes.
\newblock Unpublished manuscript.

\bibitem[\protect\citeauthoryear{Curiel}{Curiel}{2014b}]{curiel-geom-objs-nonexist-sab-uniq-efe}
Curiel, E. (2014b).
\newblock On geometric objects, the non-existence of a gravitational
  stress-energy tensor, and the uniqueness of the einstein field equation.
\newblock Submitted to \emph{Studies in History and Philosophy of Modern
  Physics}, August 2014. Latest version available online at
  \url{http://strangebeautiful.com/papers/curiel-nonexist-grav-seten-uniq-efe.pdf}.

\bibitem[\protect\citeauthoryear{Curiel}{Curiel}{2014c}]{curiel-primer-econds}
Curiel, E. (2014c).
\newblock A primer on energy conditions.
\newblock In D.~Lehmkuhl (Ed.), {\em Towards a Theory of Spacetime Theories},
  Einstein Studies, Chapter [***], pp.\  [***]. [***]: Birk\"auser.
\newblock Preprint available online at \href{http://arxiv.org/abs/1405.0403}
  {arXiv:1405.0403 [gr-qc]}.

\bibitem[\protect\citeauthoryear{Fermi}{Fermi}{1956}]{fermi-thermo}
Fermi, E. (1937[1956]).
\newblock {\em Thermodynamics}.
\newblock Dover Publications, Inc.
\newblock The Dover 1956 edition is an unabridged, unaltered republication of
  the 1937 Prentice-Hall edition.

\bibitem[\protect\citeauthoryear{Hawking}{Hawking}{1974}]{hawking-bh-explode}
Hawking, S. (1974, 01 March).
\newblock Black hole explosions?
\newblock {\em Nature\/}~{\em 248}, 30--31.
\newblock \href{http://dx.doi.org/10.1038/248030a0} {doi:10.1038/248030a0}.

\bibitem[\protect\citeauthoryear{Hawking}{Hawking}{1975}]{hawking-part-create-bh}
Hawking, S. (1975).
\newblock Particle creation by black holes.
\newblock {\em Communications in Mathematical Physics\/}~{\em 43}, 199--220.
\newblock \href{http://dx.doi.org/10.1007/BF02345020} {doi:10.1007/BF02345020}.

\bibitem[\protect\citeauthoryear{Hayward}{Hayward}{1994}]{hayward-genl-laws-bhdyns}
Hayward, S. (1994).
\newblock General laws of black hole dynamics.
\newblock {\em Physical Review D\/}~{\em 49\/}(12), 6467--6474.
\newblock \href{http://dx.doi.org/10.1103/PhysRevD.49.6467}
  {doi:10.1103/PhysRevD.49.6467}. Preprint available at
  \href{http://arxiv.org/abs/gr-qc/9303006} {arXiv:gr-qc/9303006v3}.

\bibitem[\protect\citeauthoryear{Hollands and Wald}{Hollands and
  Wald}{2012}]{hollands-wald-stab-bhs-black-brns}
Hollands, S. and R.~Wald (2012).
\newblock Stability of black holes and black branes.
\newblock \href{http://arxiv.org/abs/1201.0463} {arXiv:1201.0463v4 [gr-qc]}.

\bibitem[\protect\citeauthoryear{Israel}{Israel}{1986}]{israel-3rdlaw-bh}
Israel, W. (1986).
\newblock {T}hird {L}aw of black hole mechanics: {A} formulation of a proof.
\newblock {\em Physical Review Letters\/}~{\em 57\/}(4), 397--399.
\newblock \href{http://dx.doi.org/10.1103/PhysRevLett.57.397}
  {doi:10.1103/PhysRevLett.57.397}.

\bibitem[\protect\citeauthoryear{Jacobson}{Jacobson}{1995}]{jacobson-thermo-st-einstein-eqn-ste}
Jacobson, T. (1995, 14 August).
\newblock Thermodynamics of spacetime: {T}he {E}instein equation of state.
\newblock {\em Physical Review Letters\/}~{\em 75\/}(7), 1260--1263.
\newblock \href{http://dx.doi.org/10.1103/PhysRevLett.75.1260}
  {doi:10.1103/PhysRevLett.75.1260}. Preprint available at
  \href{http://arxiv.org/abs/gr-qc/9504004} {arXiv:gr-qc/9504004v2}.

\bibitem[\protect\citeauthoryear{Jacobson}{Jacobson}{2003}]{jacobson-intro-qft-cst}
Jacobson, T. (2003).
\newblock Introduction to quantum fields in curved spacetime and the {H}awking
  effect.
\newblock \href{http://xxx.lanl.gov/abs/gr-qc/0308048} {arXiv:gr-qc/0308048v3}.

\bibitem[\protect\citeauthoryear{Lysov and Strominger}{Lysov and
  Strominger}{2011}]{lysov-strominger-einst-navst}
Lysov, V. and A.~Strominger (2011).
\newblock From {P}etrov-{E}instein to {N}avier-{S}tokes.
\newblock \href{http://arxiv.org/absolute/hep-th/1104.5502}
  {arXiv:hep-th/1104.5502v2}.

\bibitem[\protect\citeauthoryear{Manko and Ruiz}{Manko and
  Ruiz}{2001}]{manko-ruiz-exact-soln-dbl-kerr-equil}
Manko, V. and E.~Ruiz (2001).
\newblock Exact solution of the double-{K}err equilibrium problem.
\newblock {\em Classical and Quantum Gravity\/}~{\em 18}, L11--L15.
\newblock \href{http://dx.doi.org/10.1088/0264-9381/18/2/102}
  {doi:10.1088/0264-9381/18/2/102}.

\bibitem[\protect\citeauthoryear{Maxwell}{Maxwell}{1888}]{maxwell-theory-heat-1888}
Maxwell, J.~C. (1888).
\newblock {\em Theory of Heat}.
\newblock Mineola, NY: Dover Publications, Inc.
\newblock The Dover edition of 2001 republishes in unabridged form the ninth
  edition of 1888 published by Longmans, Green and Co., London, and also
  includes the corrections and notes of Lord Rayleigh incorporated into the
  edition of 1891.

\bibitem[\protect\citeauthoryear{Penrose}{Penrose}{1979}]{penrose-sing-time-asym}
Penrose, R. (1979).
\newblock Singularities and time-asymmetry.
\newblock In S.~Hawking and W.~Israel (Eds.), {\em General Relativity: An
  {E}instein Centenary Survey}, pp.\  581--638. Cambridge University Press.

\bibitem[\protect\citeauthoryear{Planck}{Planck}{1926}]{planck97/26}
Planck, M. (1926).
\newblock {\em Thermodynamics}.
\newblock Dover Publications, Inc.
\newblock The Dover reprint of the third English edition of 1926, translated by
  A.~Ogg from the 7th German edition of 1922.

\bibitem[\protect\citeauthoryear{Schr\"{o}dinger}{Schr\"{o}dinger}{1960}]{schrodinger-stat-therm}
Schr\"{o}dinger, E. (1960).
\newblock {\em Statistical Thermodynamics\/} (Second ed.).
\newblock Cambridge: Cambridge University Press.
\newblock A course of seminar lectures, delivered in January--March 1944, at
  the School of Theoretical Physics, Dublin Institute for Advanced Studies.

\bibitem[\protect\citeauthoryear{Sciama}{Sciama}{1976}]{sciama-bhs-thermo}
Sciama, D. (1976).
\newblock Black holes and their thermodynamics.
\newblock {\em Vistas in Astronomy\/}~{\em 19, Part 4}, 385--41.
\newblock \href{http://dx.doi.org/10.1016/0083-6656(76)90052-0}
  {doi:10.1016/0083-6656(76)90052-0}.

\bibitem[\protect\citeauthoryear{Sommerfeld}{Sommerfeld}{1964}]{sommerfeld-thermo}
Sommerfeld, A. (1964).
\newblock {\em Thermodynamics and Statistical Mechanics}, Volume \textsc{v} of
  {\em Lectures on Theoretical Physics}.
\newblock New York: Academic Press.
\newblock Trans. J. Kestin. Edited and posthumously completed by F. Bopp and J.
  Meixner.

\bibitem[\protect\citeauthoryear{Thorne, Price, and MacDonald}{Thorne
  et~al.}{1986}]{thorne-etal-bhs-membrane}
Thorne, K., R.~Price, and D.~MacDonald (Eds.) (1986).
\newblock {\em Black Holes: The Membrane Paradigm}.
\newblock New Haven, CT: Yale University Press.

\bibitem[\protect\citeauthoryear{Tolman}{Tolman}{1934}]{tolman-rel-thermo-cosmo}
Tolman, R. (1934).
\newblock {\em Relativity, Thermodynamics and Cosmology}.
\newblock New York City: Dover Publications, Inc.
\newblock A 1987 facsimile of the edition published by the Oxford University
  Press, at Oxford, 1934, as part of the International Series of Monographs on
  Physics.

\bibitem[\protect\citeauthoryear{Unruh and Wald}{Unruh and
  Wald}{1982}]{unruh-wald-acc-rad-gsl}
Unruh, W. and R.~Wald (1982).
\newblock Acceleration radiation and the {G}eneralized {S}econd {L}aw of
  thermodynamics.
\newblock {\em Physical Review D\/}~{\em 25\/}(4), 942--958.
\newblock \href{http://dx.doi.org/10.1103/PhysRevD.25.942}
  {doi:10.1103/PhysRevD.25.942}.

\bibitem[\protect\citeauthoryear{Wald}{Wald}{1984}]{wald-gr}
Wald, R. (1984).
\newblock {\em General Relativity}.
\newblock Chicago: University of Chicago Press.

\bibitem[\protect\citeauthoryear{Wald}{Wald}{1994}]{wald-qft-cst}
Wald, R. (1994).
\newblock {\em Quantum Field Theory in Curved Spacetime and Black Hole
  Thermodynamics}.
\newblock Chicago: University of Chicago Press.

\bibitem[\protect\citeauthoryear{Wald}{Wald}{1997}]{wald-nernst-bh-thermo}
Wald, R. (1997).
\newblock `{N}ernst theorem' and black hole thermodynamics.
\newblock {\em Physical Review D\/}~{\em 56\/}(10), 6467--6474.
\newblock \href{http://dx.doi.org/10.1103/PhysRevD.56.6467}
  {doi:10.1103/PhysRevD.56.6467}. Preprint available at
  \href{http://xxx.lanl.gov/abs/gr-qc/9704008} {arXiv:gr-qc/9704008}.

\bibitem[\protect\citeauthoryear{Wald}{Wald}{1999}]{wald99}
Wald, R. (1999).
\newblock Gravitation, thermodynamics and quantum theory.
\newblock \href{http://xxx.lanl.gov/abs/gr-qc/9901033} {arXiv:gr-qc/9901033}.

\bibitem[\protect\citeauthoryear{Wald and Gao}{Wald and
  Gao}{2001}]{wald-gao-proc-1st-genl-2nd-charged-rot-bhs}
Wald, R. and S.~Gao (2001).
\newblock {``Physical} process version'' of the {F}irst {L}aw and the
  {G}eneralized {S}econd {L}aw for charged and rotating black holes.
\newblock {\em Physical Review D\/}~{\em 64\/}(8), 084020.
\newblock \href{http://dx.doi.org/10.1103/PhysRevD.64.084020}
  {doi:10.1103/PhysRevD.64.084020}.

\end{thebibliography}
\end{document}